\documentclass[10pt]{iopart}
\usepackage{subfigure}
\usepackage{amssymb}
\usepackage{breqn}
\usepackage[pdftex]{graphicx}   
\usepackage{epstopdf}   
\usepackage{color}
\usepackage[absolute,overlay]{textpos}
\usepackage{cancel}
\usepackage{textcomp}
\usepackage{lmodern}
\usepackage[T1]{fontenc}
\usepackage{bm}
\usepackage{siunitx}

\begin{document}

\title[Control of locked mode phase and rotation with application to modulated ECCD]{Feedforward and feedback control of locked mode phase and rotation in DIII-D with application to modulated ECCD experiments}
\author{W. Choi\(^1\),
R.J. La Haye\(^2\),
M.J. Lanctot\(^{2,a}\),
K.E.J. Olofsson\(^{1,b}\),
E.J. Strait\(^2\),
R. Sweeney\(^{1,c}\),
F. A.~Volpe\(^1\) and The DIII-D Team\(^2\)}

\address{\(^1\) Columbia University, New York, NY 10027}
\address{\(^2\) General Atomics, San Diego, CA 92121}
\address{\(^a\) Present address: U.S. Department of Energy, SW Washington, DC 20585} 
\address{\(^{b}\) Present address: General Atomics, San Diego, CA 92121}
\address{\(^{c}\) Present address: ITER Organization, 13067 St Paul Lez Durance, France} 

\begin{abstract}
The toroidal phase and rotation of otherwise locked magnetic islands of toroidal mode number \(n\)=1 are controlled in the DIII-D tokamak by means of applied magnetic perturbations of \(n\)=1. 
Pre-emptive perturbations were applied in feedforward to "catch" the mode as it slowed down and entrain it to the rotating field before complete locking, thus avoiding the associated major confinement degradation. 
Additionally, for the first time, the phase of the perturbation was optimized in real-time, in feedback with magnetic measurements, in order for the mode's phase to closely match a prescribed phase, as a function of time. 
Experimental results confirm the capability to hold the mode in a given fixed-phase or to rotate it at up to 20 Hz with good uniformity. 
The control-coil currents utilized in the experiments agree with the requirements estimated by an electromechanical model. 
Moreover, controlled rotation at 20 Hz was combined with Electron Cyclotron Current Drive (ECCD) modulated at the same frequency. 
This is simpler than regulating the ECCD modulation in feedback with spontaneous mode rotation, and enables repetitive, reproducible ECCD deposition at or near the island O-point, X-point and locations in between, for careful studies of how this affects the island stability. 
Current drive was found to be radially misaligned relative to the island, and resulting growth and shrinkage of islands matched expectations of the Modified Rutherford Equation for some discharges presented here.
Finally, simulations predict the as designed ITER 3D coils can entrain a small island at sub-10~Hz frequencies.

\end{abstract}

\ioptwocol
\section{Introduction}
Current perturbations, in the form of a deficit in pressure-driven bootstrap current, can cause magnetic reconnection and formation of Neoclassical Tearing Modes (NTMs) in the plasma \cite{lahaye2006}. 
Islands\textemdash identified by poloidal and toroidal mode numbers \(m\) and \(n\) respectively, 
form at "rational" surfaces, where the safety factor equals \(q=m/n\)\textemdash 
can cause a non-axisymmetric  local flattening in the pressure profile, 
which increases the deficit in bootstrap current and drives the island to grow larger.

While initially rotating at the plasma's natural rotation frequency, these non-axisymmetric current filaments can induce eddy currents in the vacuum vessel wall, which in turn drag on the NTM, causing it to slow and eventually lock to residual error fields \cite{nave1990}. 
These locked modes (LMs) can quickly grow in amplitude, deteriorating plasma confinement by deforming pressure profile as in \cite{chang1990} and, in the worst case scenario, cause a major disruption \cite{morris1992}.

A number of approaches have been studied to mitigate the deleterious effects of locked modes, including: avoid seeding the initial mode \cite{lahaye2000}, suppression during the rotating phase \cite{lahaye2006}, or suppression of the locked mode itself \cite{volpe2015}.
It has been demonstrated that islands can be suppressed by replacing the deficit current \cite{gantenbein2000, lahaye2006, hennen2010},  using electron cyclotron current drive (ECCD) \cite{prater2004}, given proper alignment. 
In a system with a fixed launcher location, the deposition region is constrained in major radius by resonance location, in the toroidal coordinate by purpose (current drive or heating-only), and in height by the flux surface on which power is desired.
Thus the intersection of the LM O-point with the deposition region can only be accomplished by controlling the phase of the mode.

\begin{figure}[t]
       \includegraphics[scale=0.6,trim={0 0 0.2cm 0},clip]{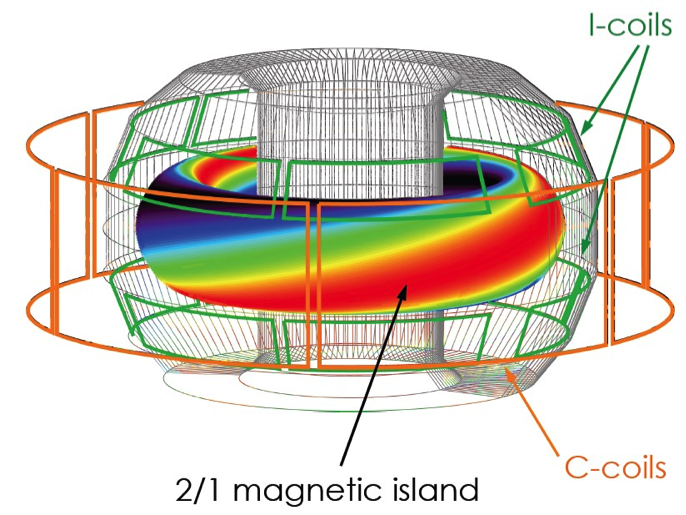}
       \caption{Perturbed current pattern, sinusoidal in helical angle \(m\theta - n\phi\), of a \(m/n=2/1\) magnetic island mapped onto a thin surface.
		Also depicted are the internal I-coils (green) and external C-coils (red) used to generate 3D fields on DIII-D.} 
\label{fig:islandExample}
\end{figure}

One common technique for controlling the mode phase is to apply resonant magnetic perturbations (RMPs) by means of non-axisymmetric coils. 
They can be either internal or external to the vessel, like the I- and C-coils at DIII-D, respectively, shown in figure \ref{fig:islandExample}. 
As predicted by theory \cite{fitzpatrick1991}, such coils can be used to apply an electromagnetic torque to the mode and thus control its toroidal phase. 
The simplest usage would be to apply a pre-determined RMP, either static or rotating, with the expectation that the mode will align to it or, more precisely, to the resultant of the applied RMP and error field.
This feedforward method has previously been shown to entrain modes at DITE \cite{morris1990}, TEXTOR \cite{liang2007}, and DIII-D \cite{volpe2015}. 
Experiments at TEXTOR were conducted on tearing modes driven by error field penetration (where the error field in question was the RMP itself) \cite{koslowski2006}. Experiments at DIII-D were carried out on pressure-driven NTMs \cite{volpe2015,volpe2009,shiraki2014}.
Other methods include using a feedback algorithm where the measured mode amplitude is multiplied by a complex gain to obtain the applied RMP \cite{okabayashi2014}. At DIII-D, tuning the complex gain yielded control of the mode rotation frequency up to 50~Hz \cite{okabayashi2017}, while feedforward reached rotation frequencies up to 300~Hz \cite{volpe2014}. 

Both the feedforward and feedback approaches described forced the mode to rotate despite drag from the wall. 
However, neither approach had fine control of the mode {\em phase}. 
Such a feature is highly desirable when controlled mode rotation is combined with modulated ECCD, in order to suppress the mode and restore good confinement. 
This is because mode rotation alone, without ECCD, would only be stabilizing (via rotation shear and wall shielding) if very fast, 
well above the inverse wall-time \cite{devries1996}. 
Otherwise, ECCD is necessary to stabilize the rotating mode. 
Preferably, the ECCD should be modulated in phase with the transit of the island O-point in the deposition region. 
This poses the need to either measure the island phase and adjust the ECCD modulation accordingly, in real time \cite{maraschek2007} or, as shown in the present work, to prescribe the island phase in advance, as a function of time, and pre-program the ECCD modulation accordingly.

The theory of a feedback controller of the island phase was laid out in earlier work \cite{olofsson2016}.  
Assuming a saturated island of fixed width, its toroidal rotation was modeled under various conditions and compared to experimental observations. 
Using this representation, the system's response to an arbitrary control scheme was tested and optimized. 
A modified control scheme was ultimately implemented, the results of which are presented in section \ref{sec:implementController} of the present paper. 

The paper is organized as follows: section \ref{sec:preemptive} shows a new technique of preemptive entrainment in feedforward. 
Section \ref{sec:implementController} describes the feedback phase controller as implemented on DIII-D. 
Section \ref{sec:experimentalResults} presents the experimental results. 
Section \ref{sec:ITERsimulation} then extends the model to predict entrainment capabilities for ITER.

\section{Feedforward preemptive entrainment}
\label{sec:preemptive}

A novel technique of preemptive entrainment is studied here, where a rotating RMP is applied early in the current flattop period, before the mode decelerates significantly and locks. 
When the \(m/n\)=2/1 rotating precursor forms, initially in the kHz range, and decelerates due to wall drag, the mode is expected to lock onto the existing rotating RMP instead of locking to the wall. 
In theory, the prevention of complete mode locking would limit the island growth, as it is partially stabilized by rotation in the presence of a conducting wall \cite{brennan2014} and by rotation shear \cite{chen1990, buttery2008, lahaye2009}.

\subsection{Experimental result}
\begin{figure}[h]
\centering
\includegraphics[scale=0.9,trim={0cm 0.2cm 0 0cm},clip]{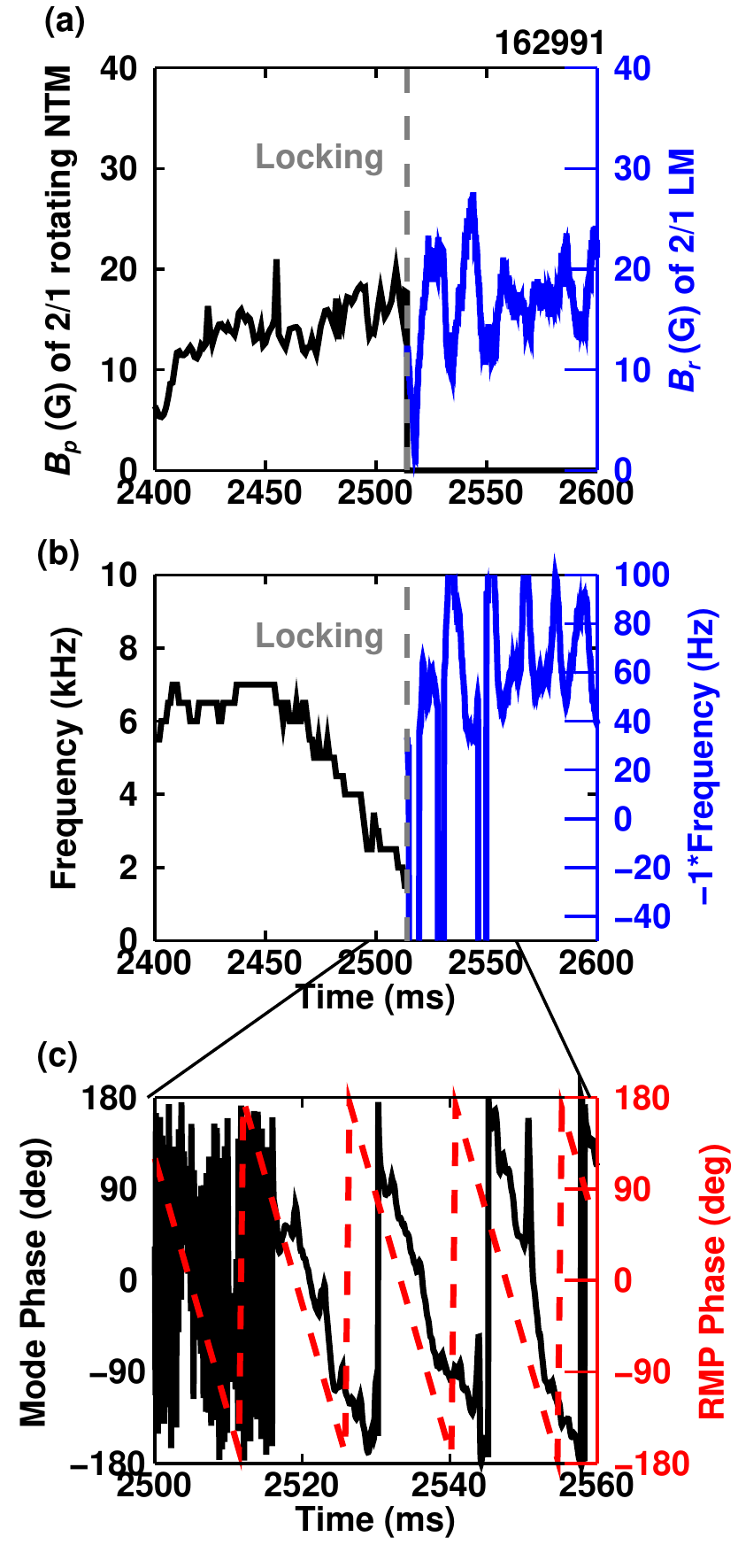}
\caption{(a) Amplitude, (b) frequency, and (c) phase of the NTM, before and after locking. Note that instead of dropping to zero, the frequency after locking is around 70~Hz (with some noise), showing the island (black) is locked to the applied RMP (red) rotation frequency. Here, negative frequency is in the plasma rotation direction.}
\label{fig:preemptiveEntrainment}
\end{figure}

An experiment was performed with the preemptive entrainment technique at DIII-D. 
The discharges had a flat top plasma current of 0.9 to 1.0~MA and a toroidal field of 1.7~T, 
giving the plasma a safety factor \(q_{95}\) between 4.8 to 5.8 during the flat top.
After current flat top was reached at 700~ms,  a RMP rotating at 70~Hz was applied starting at 1300~ms, with 4.3~kA of current in the internal I-coils. 
An average of 3~MW of neutral beam power pushed the plasma into H-mode at approximately 1760~ms, after which NBI power was increased up to a peak of 9~MW.
In the discharge shown in figure \ref{fig:preemptiveEntrainment}, the 2/1 NTM appeared at around 2392~ms, initially rotating at about 7 kHz. 
Neutral beam power was reduced to roughly 5.8~MW near this time, which provided approximately 0.4~Nm of torque to the plasma for the rest of the discharge.

As the island rotation was reduced to near zero at 2514~ms, it locked to the 70~Hz rotating RMP without locking to the intrinsic residual DC error field. 
To be more precise, the island tends to align to the resultant of the rotating RMP and static error field. 
This explains why its rotation is non-uniform (figure \ref{fig:preemptiveEntrainment}(c)) and why the frequency oscillates around 70~Hz  on a sub-period timescale (figure \ref{fig:preemptiveEntrainment}(b)). 
Note that the {\em instantaneous} rotation frequency plotted in figure \ref{fig:preemptiveEntrainment}(b) is defined as the time-derivative of the toroidal phase signal in figure \ref{fig:preemptiveEntrainment}(c), where negative frequency is in the direction of natural plasma rotation. 
Also, the frequency oscillations can be so large that the mode rotation can temporarily change direction (figure \ref{fig:preemptiveEntrainment}(c)) \cite{volpe2014}, and the rotation frequency can temporarily vanish, for a small fraction of the 14.3~ms rotation period. 
Four instances of vanishing rotation frequency can be noticed in figure \ref{fig:preemptiveEntrainment}(b), but they should not be interpreted as mode locking.  
Other shots from the same experiment exhibit similar behaviour.

The measured amplitude and phase of the mode were corroborated between measurements of the perturbed poloidal and radial fields, as well as using the electron cyclotron emission diagnostic \cite{austin2003}.
Here and in the remainder of the paper, mode phase is defined as the toroidal position of peak perturbed radial magnetic field as measured at the outboard mid-plane.
The poloidal field measurements come from an \(n=1\) fit of a toroidal array of Mirnov probes, located inside the vessel.
The perturbed radial field was measured by external saddle loops differenced (ESLD), where sensors 180\(^\circ\) apart are subtracted to obtain odd-\(n\) signals.
Further discussion of the magnetic sensors on DIII-D and their capabilities to measure 3D phenomenon can be found in \cite{strait2006, king2014}.
As the applied RMP and the mode rotate at the same frequency (except for the oscillations just discussed), it was important to isolate the actual mode measurement and remove any coil-sensor pickup of the applied fields. 
This was realized by a.c. compensation \cite{hanson2016}. 

In summary, this preemptive entrainment technique traded complete mode locking for a slowly rotating entrained island. 
This technique requires a rough prediction of when the mode might appear, and with what amplitude, so that the rotating RMP is only applied when needed, and with an amplitude just sufficient for entrainment, so that its negative impact on confinement can be minimized. 

\subsection{Discussion and simulation}
By definition, a preemptive rotating RMP is applied before the rotating mode has appeared. 
Therefore, preemptive entrainment can only be applied in feedforward as there is no signal to feed back on.

\begin{figure}[t]
\begin{flushright}
\textbf{\scriptsize{162991}}
\vspace{-0.35cm}
\end{flushright}
\centering
\hspace{-0.3cm}\includegraphics[scale=1,trim={0cm 0cm 0cm 0cm},clip]{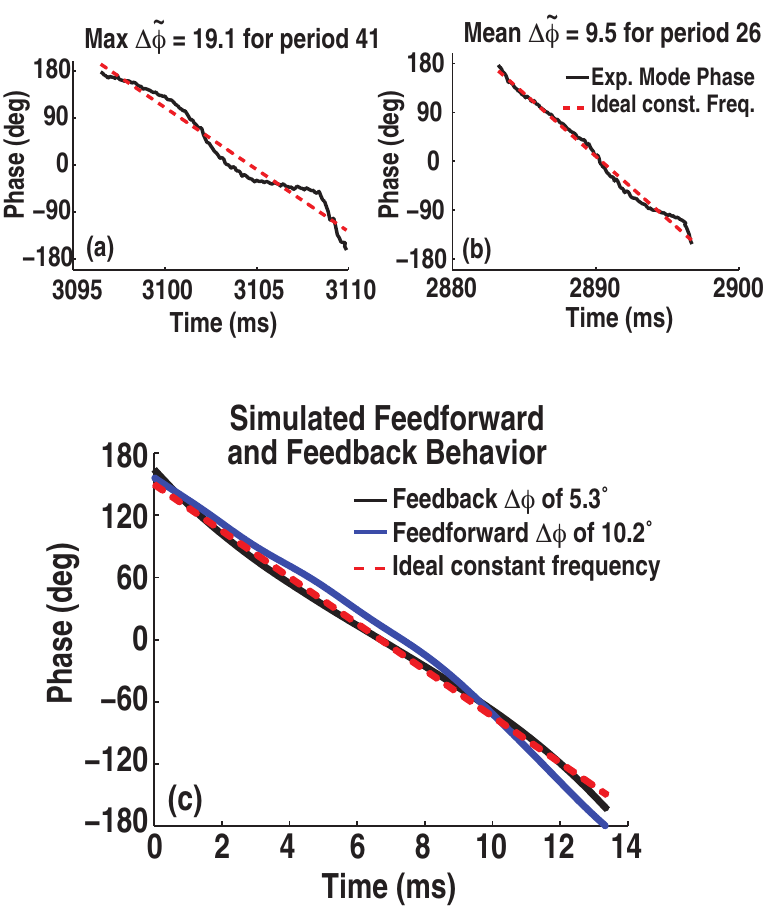}
\caption{Nonuniform island rotation (black, solid) in response to an applied RMP rotating uniformly (red, dashed), within a single period. (a)-(b) Measurements in the case of maximum and mean nonuniformity, respectively, as measured by  \(\Delta\tilde{\Phi}\). (c) Simulation illustrating reduced nonuniformity when switching from a feedforward scheme (blue) to feedback control (black).}
\label{fig:nonuniformRotation}
\end{figure}

Even if the applied fields rotate uniformly, the island will very likely rotate non-uniformly, 
due to electromagnetic torques from residual error fields, fluctuating island widths, and changing plasma conditions.
An extreme case and a typical example are illustrated in figures \ref{fig:nonuniformRotation}(a) and \ref{fig:nonuniformRotation}(b), respectively. 
Both cases are taken from the same experiment reported in figure \ref{fig:preemptiveEntrainment}, although at different times. 
Here the amount of deviation from uniform rotation is quantified by the root-mean-square of the difference between the actual mode phase and a fixed-frequency trajectory. 
These r.m.s. phase-differences are evaluated over one period, and are denoted by \(\Delta\tilde{\Phi}\). 
For the 76 periods in which the mode was entrained, the values of \(\Delta\tilde{\Phi}\) ranged between 3.5\(^\circ\) and 19\(^\circ\) (figure \ref{fig:nonuniformRotation}(a)), with a mean of 9.5\(^\circ\) (figure \ref{fig:nonuniformRotation}(b)). 

It is interesting to note that, in the experiment, the phase deviation appears to be a mix of \(n=1\) and \(n=2\) perturbations, exemplified by the asymmetric deviations of the measured phase in black from the ideal red dashed trace.  
As both the mode and applied RMP are \(n=1\), this behaviour might be the result of an \(n=2\) error field interacting with the mode.

The dynamics of the island are governed by the equation of motion:
\begin{equation}
I\frac{d^2\phi}{dt^2} = T_{wall}+T_{EF}+T_{RMP}
\label{eqn:simpleEoM}
\end{equation}
where \(I\) and \(\phi\) are the moment of inertia and toroidal phase of the island, respectively. 
On the right-hand-side of the equation are the electromagnetic torques exerted by the wall, static error field, and applied RMP onto the island.
These torques are calculated from a numerical model \cite{olofsson2016} which treats the island as a single \(m/n=2/1\) Fourier mode in cylindrical geometry,
and the perturbed or induced currents and applied or intrinsic magnetic fields are represented by phasors.
This is the equation used in time-dependent simulations herein.

A few simplifying assumptions have been made: no other tearing modes exist or can interact with the 2/1 mode of interest, the plasma around the island has low rotation and is not imparting momentum on the island, and torque from the neutral beams on the island is negligible. 
This last assumption is valid for either balanced co- and counter-injection, or for beam deposition away from the island location. 
In this experiment, the NBI torque deposited directly into the island, estimated based on ratio of the poloidal cross-sectional areas of the island and the plasma, is approximately 0.04~Nm, which can be neglected when compared to electromagnetic torques on the order of 1~Nm.

In simulations of the preemptive feedforward entrainment, an uncorrected \(n=1\) error field of 0.5~G at the rational surface causes an \(n=1\) perturbation to the rotation with an average phase deviation of \(\Delta\tilde{\Phi}=10.2^\circ\). 
This error field amplitude was chosen to approximately match the experimental mean value of \(\Delta\tilde{\Phi}=9.5^\circ\) in simulation. 
Simulations performed under the same plasma conditions and for the same error field indicate that the deviation is reduced to \(\Delta\tilde{\Phi}=5.3^\circ\) when feedback is used (figure \ref{fig:nonuniformRotation}(c)), which is described in the next section.

\section{Implementation of feedback phase controller}
\label{sec:implementController}
\begin{figure}[t]
\includegraphics[scale=1,trim={0cm 0cm 0 0cm},clip]{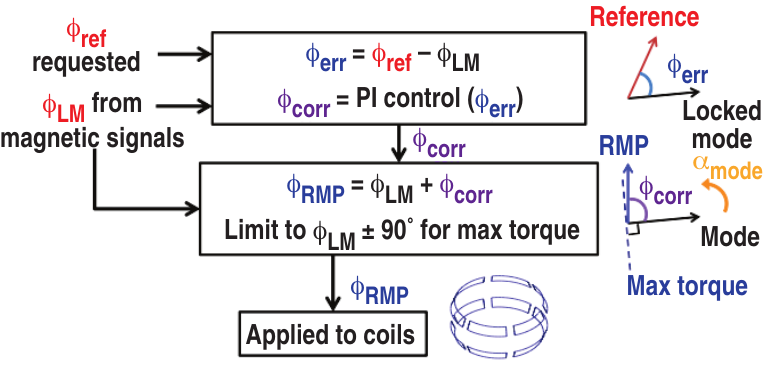}
\caption{A schematic diagram of the proportional-integral controller implemented on DIII-D.}
\label{fig:controllerFlowChart}
\end{figure}

A feedback controller of the 2/1 mode phase was implemented in the DIII-D plasma control system, as shown in figure \ref{fig:controllerFlowChart}. 
Using the external saddle loops differenced (ESLD) signal as input, the algorithm calculates the amplitude and phase of the \(n=1\) locked or slowly rotating mode. 
The issue of drifts often associated with time-integrated signals is addressed by subtracting a baseline from each measured value. 
Such a baseline is evaluated 50~ms before mode locking, when the mode is still rotating rapidly (faster than 100~Hz) and, consequently, is not detectable by the ESLDs, due to wall shielding.

The operation of the controller is described here.
The calculated phase of the island \(\phi_{LM}\) is compared to a pre-determined reference phase \(\phi_{ref}\). 
The discrepancy or error angle between the two, \(\phi_{err}\), is used as input to a proportional-integral (PI) controller. 
Note that the subscript \(_{err}\) is not related to (and should not be confused with) the error field. 
The PI controller accounts for the present error angle as well as its time history with:
\begin{equation}
\phi_{corr}(t) = K_p \phi_{err}(t) + K_i \int_{t_0}^t \phi_{err}(t') dt'
\end{equation}
where \(K_p\) and \(K_i\) are the proportional and integral gains respectively, and \(t_0\) is the time at which the controller is turned on.
The output of the optimal correction angle \(\phi_{corr}\) advances the phase of the applied perturbation, \(\phi_{RMP}\), with respect to \(\phi_{LM}\).
This correction is then clipped to \(\phi_{LM}\pm90^\circ\) for maximum torque.  
Ultimately, a.c. currents of constant amplitude are delivered to the internal 3D field coils (I-coils) in order to apply an \(n\)=1 RMP at the desired phase \(\phi_{RMP}(t)\). 
This algorithm was implemented on a CPU with a cycle time of 50~\(\mu\)s, much shorter than the wall time of about 3~ms. 
This fast timescale allows rapid adjustments of \(\phi_{RMP}(t)\), in order for \(\phi_{LM}(t)\) to faithfully match the slowly evolving, desired \(\phi_{ref}(t)\). 
In the experiments presented here \(\phi_{ref}(t)\) is either static or evolving at 20~Hz.

The present approach is satisfactory at frequencies well below the inverse wall-time (about 50~Hz at DIII-D). 
For faster entrainment and associated stabilizing effects, the amplitudes of the coil-currents should be increased as a function of the requested rotation frequency to overcome the effect of the increasing wall shielding.
This is left as a future improvement.

\subsection{Mode phase control in experiment}
An experiment was performed on DIII-D with the goal of using the new feedback algorithm to achieve smooth entrainment of the mode, combined with synchronized deposition of ECCD to suppress the island amplitude. 
Here "smooth" refers to uniform rotation, both within a single rotation period, as well as from period to period.

The DIII-D discharges presented in the remainder of this paper have the following parameters:
a plasma current of 1.0~MA and toroidal field of 1.7~T gave a safety factor between 4.3 and 4.5.
H-mode was obtained at approximately 1500~ms and subsequently lost around 2200~ms, when an initially rotating 2/1 NTM appears, quickly decelerates and becomes locked within 200~ms.
The locking event triggers a response in the controller and, for later shots discussed in section \ref{sec:experimentalResults}, gyrotron power.

\begin{figure}[t]
\centering
\includegraphics[scale=0.48,trim={0cm 0cm 0 0cm},clip]{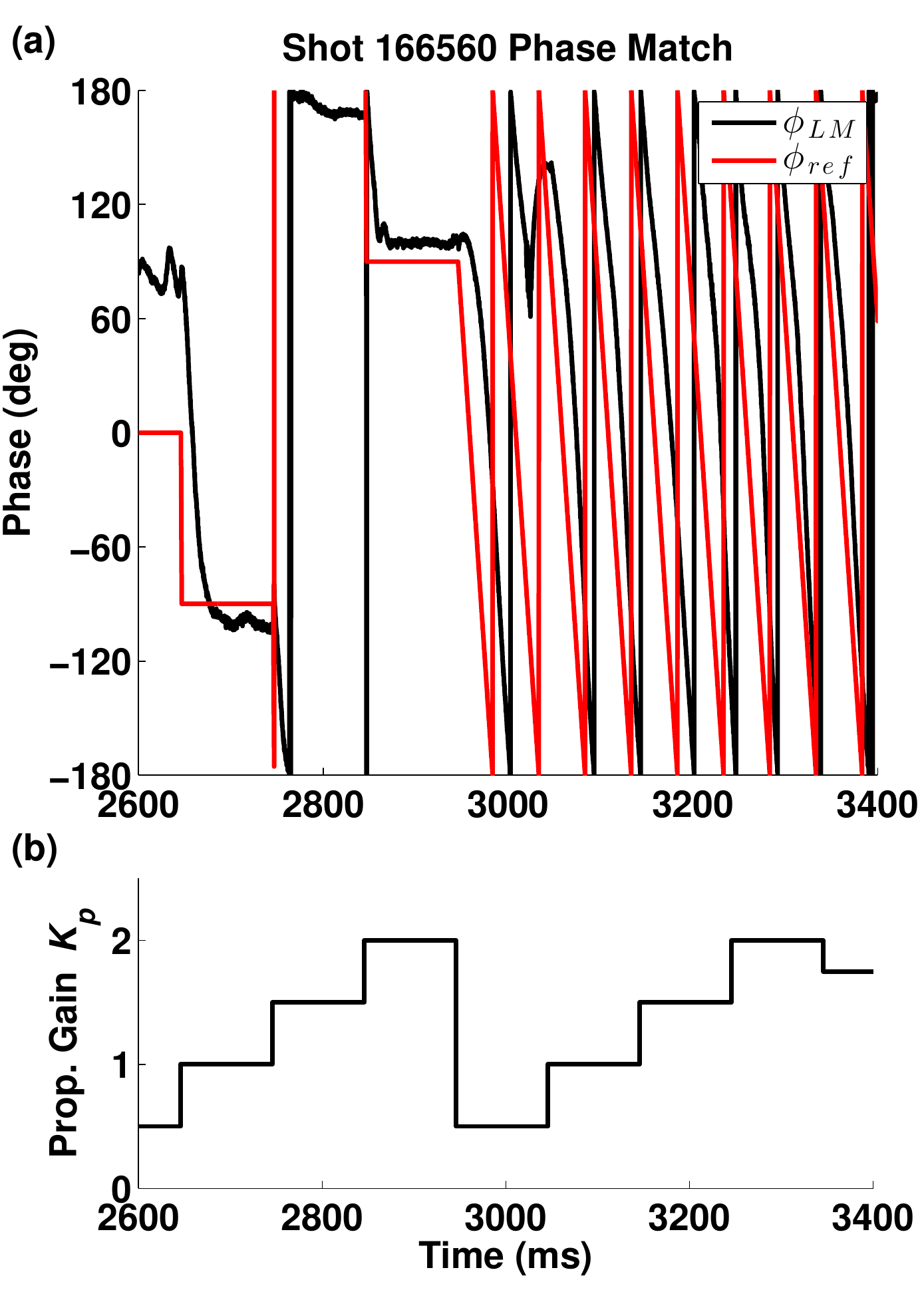}
\caption{The measured and reference phase of a locked mode, as gain is varied every 100~ms. In both fixed and rotating reference cases, higher gain results in better alignment as expected.}
\label{fig:matchModePhase}
\end{figure}

The capability of the controller was tested with a pre-programmed reference phase for fixed-phase and 20~Hz rotation. 
Figure \ref{fig:matchModePhase} shows the results of a shot where only proportional feedback was used, with the proportional gain \(K_p\) stepped over four different values in each type of request.

\begin{figure}[t]
\centering
\subfigure{
\includegraphics[scale=0.4,trim={0cm 0cm 0 0},clip]{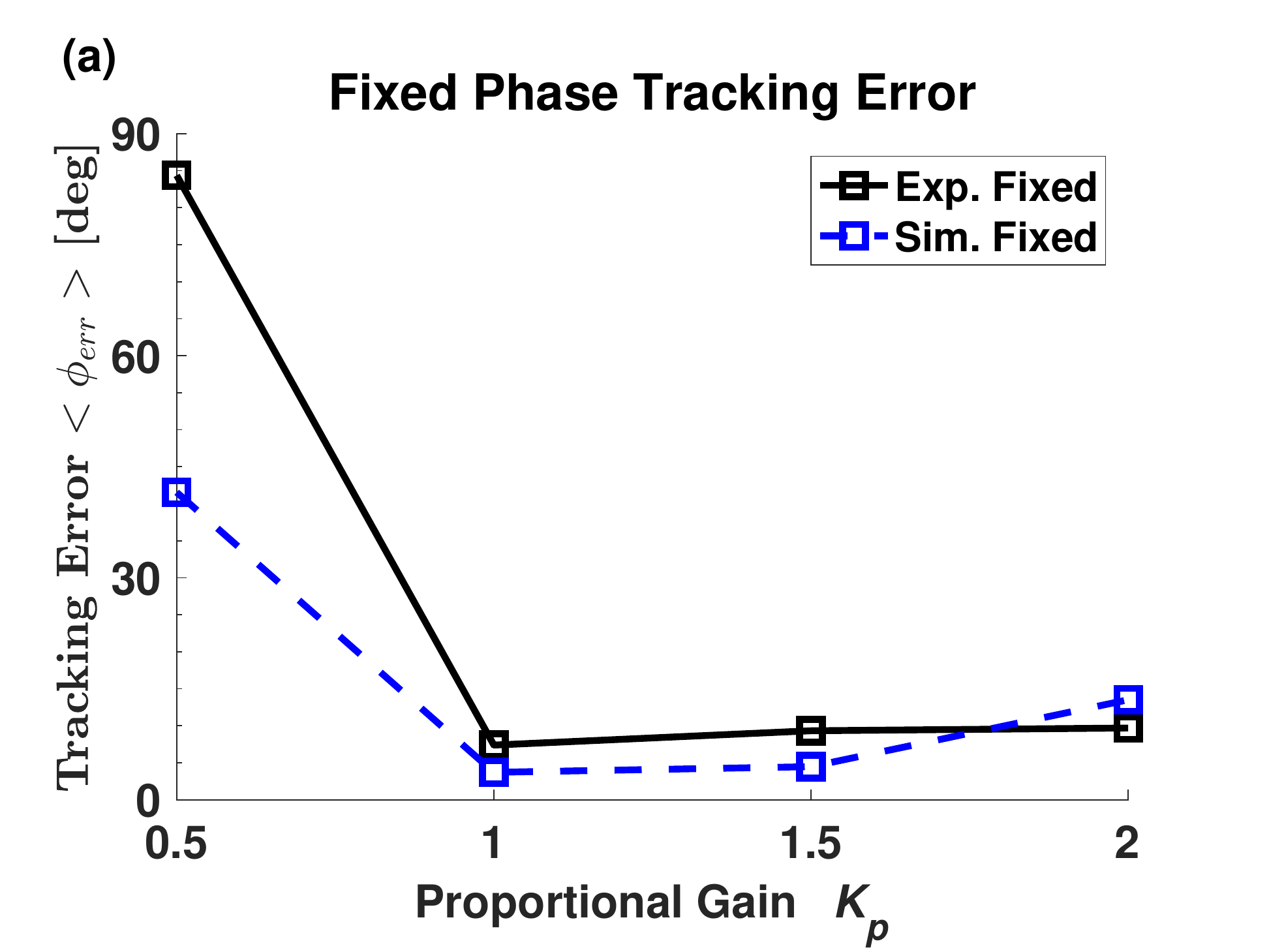}
}
\subfigure{
\includegraphics[scale=0.4,trim={0cm 0cm 0 0},clip]{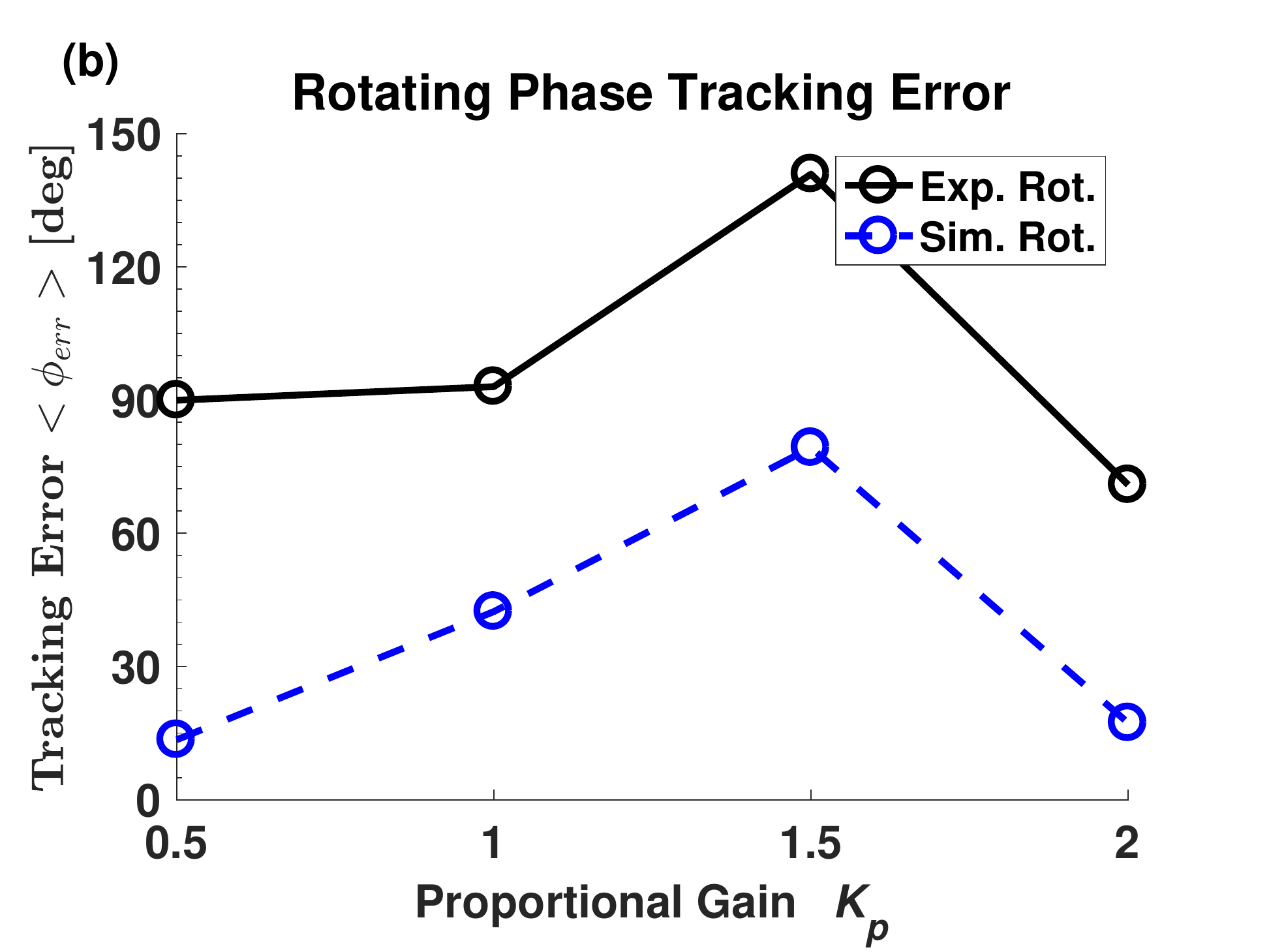}
}
\caption{Experimental and simulated tracking error (defined as discrepancy between reference phase and actual mode phase averaged over each 200~ms period) for (a) fixed phase and (b) 20~Hz rotating phase references respectively.}
\label{fig:trackingError}
\end{figure}

The agreement between the requested LM phase \(\phi_{ref}\) and the actual, magnetically measured LM phase \(\phi_{LM}\) is qualitatively evident from fig.~\ref{fig:matchModePhase}(a).
The discrepancy between the two, \(\phi_{err}\), averaged over the duration of each fixed proportional gain, is used as a quantitative figure of merit. 
This averaged discrepancy \(<\phi_{err}>\)  is calculated for both the experimental measurements (black symbols in figure \ref{fig:trackingError}) and the equivalent simulated scenario (blue symbols). 
All simulations in this section used the same parameters as in shot 166560, including the same radial position of the island, and same RMP strength, corresponding to 2.7~kA of current in the I-coils.

In the fixed-phase case (figure \ref{fig:trackingError}(a)), it is clear that the controller can control the LM phase as desired, as long as the proportional gain \(K_p\) exceeds a threshold located in the interval \(0.5<K_p<1\).

The entrainment tests (figure \ref{fig:trackingError}(b)) exhibit larger errors.
In fact, the lowest proportional gain \(K_p\)  of 0.5 in the scan was unable to entrain the mode and, at \(t\simeq 3040 ms\), it allowed the mode to "slip" backwards and catch the next rotation (figure \ref{fig:matchModePhase}(a)).
This slippage resulted in a brief period in which \(\phi_{LM}(t)\) matched \(\phi_{ref}(t)\), which may have reduced the apparent tracking error for several periods that followed (into the next proportional gain value).
Higher gain, \(K_p\ge 1\), yielded continuous entrainment without such slipping events, but with some significant discrepancy \(\phi_{err}\).
In reality, finite, relatively large discrepancies were actually expected (see blue symbols in figure \ref{fig:trackingError}(b)) as a result of how the phase-controller \cite{olofsson2016} is expected to perform at 20~Hz. 
This is due to currents induced in the wall not yet being completely accounted for in the controller.
As a matter of fact, experiments exhibited even larger tracking errors (black symbols in figure \ref{fig:trackingError}(b)), but this was also reasonable, as it is well-known that a proportional-only controller cannot track a ramped reference without error \cite{ogataControlText}.
Adjustments of controller gains achieved the desired smooth rotation and brought average \(\phi_{error}\) to within acceptable levels (\(\lesssim 30^\circ\)) during later shots.
The performance of a simple proportional-integral controller is satisfactory for this first attempt of feedback control of mode phase.
Advanced techniques such as iterative learning control \cite{felici2015, ravensbergen2017} can be adopted for further improvement.

\subsection{Simulation of proportional-only control}
Other major factors contributing to the tracking error in this proportional-only scheme are the island width and its proximity to the wall, or equivalently, the position of the \(q=2\) surface in normalized minor radius \(\rho\).

For these {\em steady-state} simulations, the effect of the error field must be dropped from equation \ref{eqn:simpleEoM}, to prevent it from adding a sub-period perturbation to an otherwise uniform  entrainment.
Experimentally, this can be achieved if the error field is well-corrected and made negligible compared with the applied, slowly rotating RMP.
The resultant motion will be a balance between the wall torque and the applied RMP torque only:

\begin{center}
\begin{equation}
	0 = T_{wall}+T_{RMP}
	\label{eqn:smoothEntrain}
\end{equation}
\end{center}

\begin{figure}[t]
\centering
\includegraphics[width=0.5\textwidth]{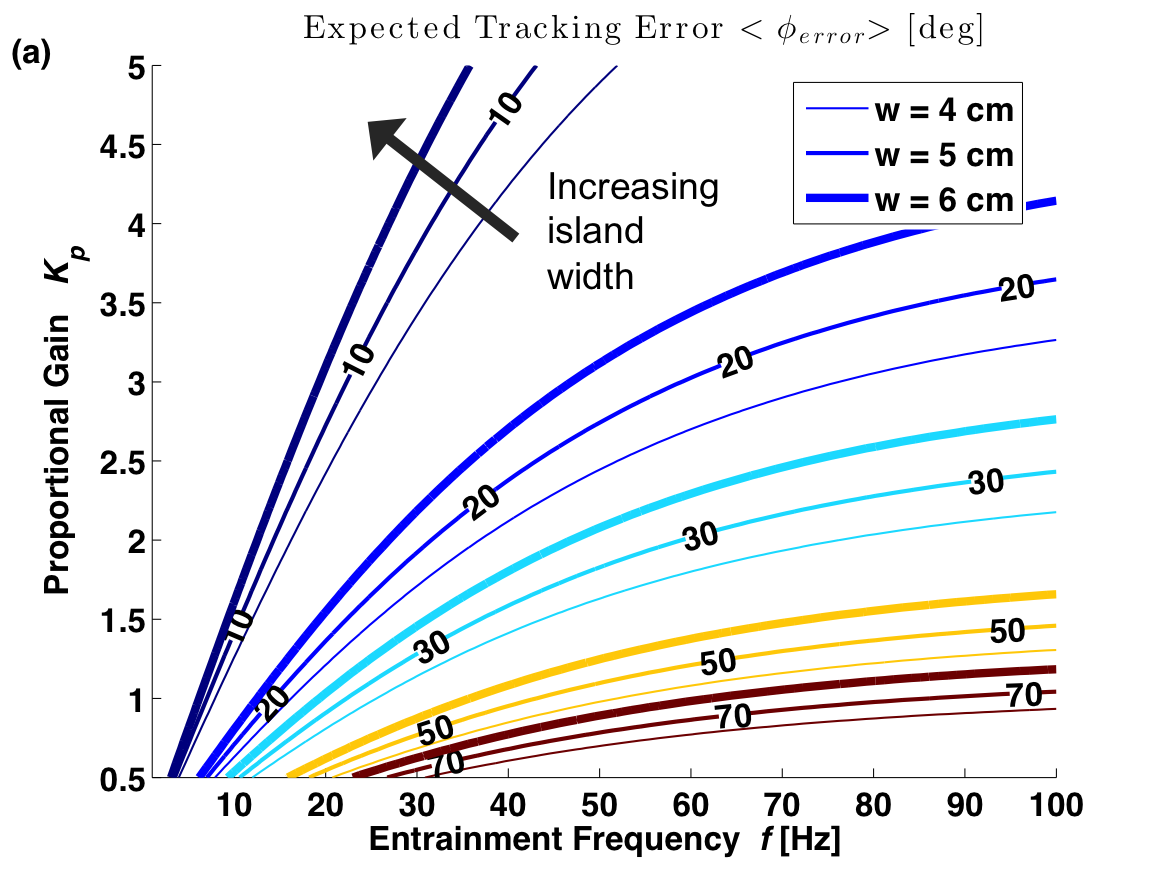}
\includegraphics[width=0.5\textwidth]{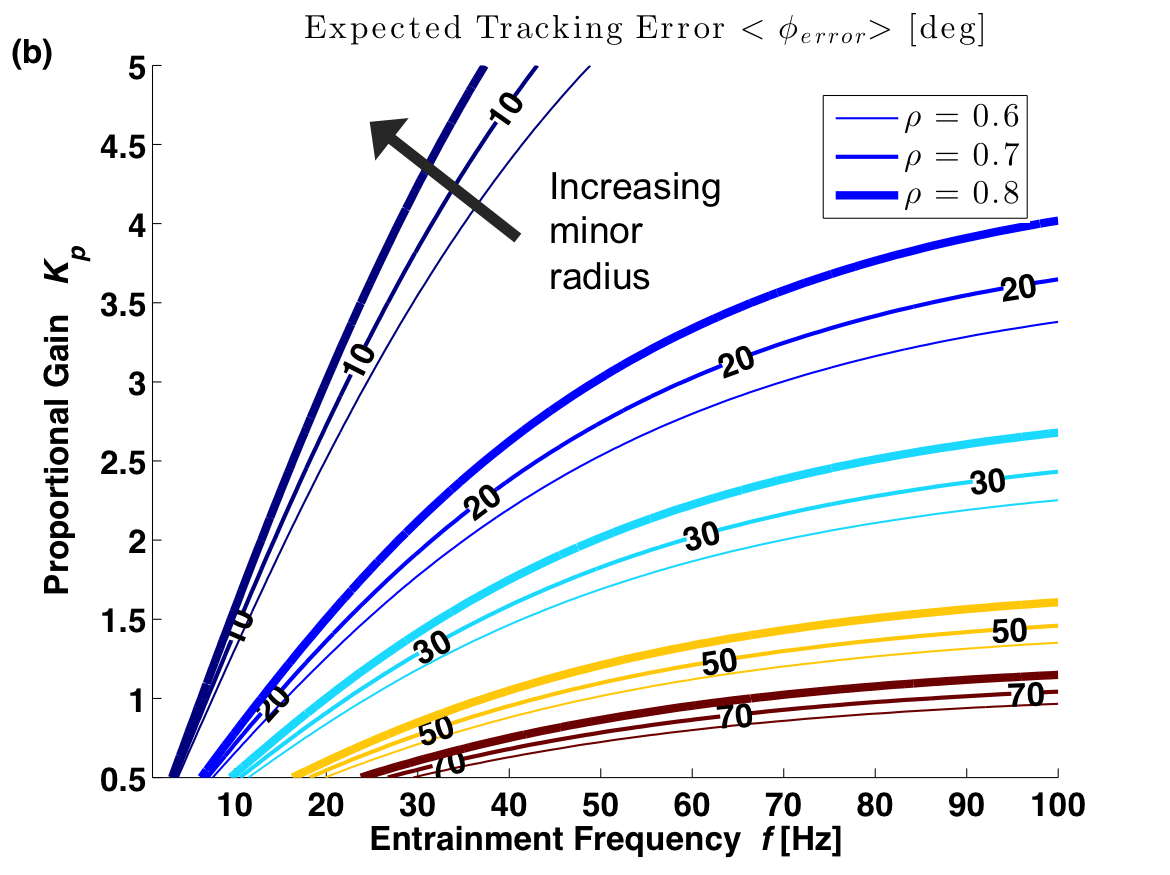}
\caption{Simulated tracking error in a proportional only controller, as dependent on the gain used, the requested frequency, as well as (a) the island width and (b) position of the \(q=2\) surface.
Three contours are presented in each panel, where the colours indicate levels of tracking error in each case and thickness differentiates the scenarios.}
\label{fig:trackingErrorContours}
\end{figure}

Figure \ref{fig:trackingErrorContours} shows the toroidal tracking errors expected from simulations, as described by equation \ref{eqn:smoothEntrain}, where the phase of the RMP is determined by the proportional control.
Each panel presents three colour contours with different thicknesses for changing (a) island width or (b) normalized radial position of the \(q=2\) surface.
For example, the contour with medium thickness lines in panel (a) shows that 
a 5~cm island entrained at 30~Hz with \(K_p\) of 2 has a predicted tracking error of 20\(^\circ\), 
whereas a 4~cm (thin lines) or 6~cm (thick lines) island using the same parameters would have errors of 17\(^\circ\) and 22\(^\circ\) respectively.
This expected lag increases for larger islands or shorter distance from the wall, due to the different dependencies of the RMP and wall torques on these parameters.
A higher entrainment frequency also increases the error due to stronger wall shielding of the applied RMP and increased wall torque (peaks at the inverse wall time) on the mode.
Simulations show that increasing the gain reduces the tracking error,  but at the risk of overall system stability in transient evolution\cite{ogataControlText}.

As it currently stands, using saddle loops {\em external} to the vessel as sensors prevents the controller from detecting modes rotating much faster than the inverse wall time. 
Even at frequencies yielding measurable signals (comparable with the inverse wall time, or approaching it), shielding presents issues: the measured mode phase lags behind the true mode phase. 
A future combination of {\em internal} sensors (whether Mirnov or saddle loops) and {\em real-time} a.c. compensation for coil-sensor pickup will contribute to deploying the controller at frequencies above the present limit.

\section{Combining mode entrainment and modulated ECCD}
\label{sec:experimentalResults}
\subsection{Experimental results}
After mode entrainment was achieved, modulated ECCD with a 50\% duty cycle was deposited in synchronization with the mode rotation at 20~Hz, corresponding to a period of 50~ms.
For comparison, typical NTM evolution timescale (\(w/(dw/dt)\)) in the present experiment is between 40 and 200~ms.
Note that in earlier experiments at DIII-D and AUG \cite{maraschek2007}, ECCD was modulated in feedback with island phase measurements, though it is a challenge to do this in real-time with available diagnostics.
More simply, here the island phase is prescribed by a feedback controller as a function of time, and the ECCD modulation is simply programmed in advance.
Average phase error of \(\lesssim 30^\circ\) was considered negligible compared to the toroidal 180\(^\circ\)  for which the ECCD was turned on.
Six gyrotrons were used in this experiment to provide up to 3.5 MW of heating and drive about 22.5~kA of current in the co-plasma current direction \cite{lohr2005}.
It should be noted that the EC current is established on an electron-electron collisional timescale, which is significantly shorter than the evolution of interest and can be considered instantaneous.
However, externally driven current effect changes in the current profile much more slowly, as discussed later.

The four plasma discharges used for analysis follow a similar trajectory in their overall discharge evolution.
After reaching current flat-top and entering H-mode, a \(m/n=2/1\) NTM appeared with a rotation frequency in the high kHz range.
The NTM was allowed to grow and decelerate without intervention.
Modulated ECCD and rotating RMPs were applied upon detection of mode locking.
From here, the measured mode amplitudes oscillated until programmed \(I_p\) ramp-down.
In contrast, an earlier reference shot with no ECCD disrupted shortly after locking.

\begin{figure*}[t]
\begin{center}
\includegraphics[width=\textwidth]{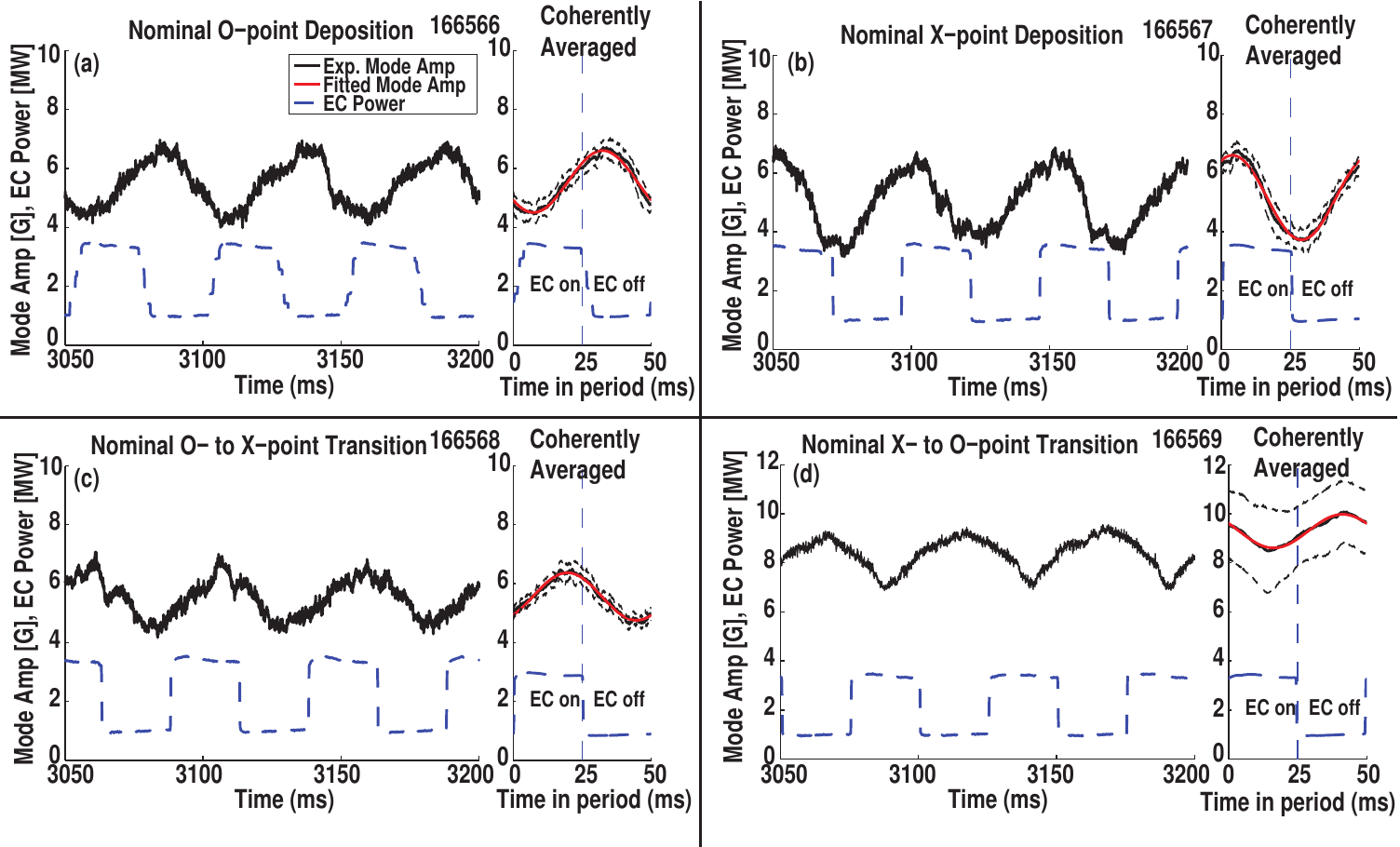}
\end{center}
\caption{Plots of measured mode amplitude, in black, and electron cyclotron power, in blue. A coherent average of the mode amplitude  over 25 periods is shown for each shot on the right, where the dotted lines mark one standard deviation away from mean black line. A sinusoidal fit to each averaged mode amplitude is shown in red. (a) Mode amplitude is increased when ECCD is thought to be deposited in the O-point, against expectation. (b) X-point deposition correlates with mode amplitude decrease.
(c) Mode appears to first grow, then be suppressed as deposition transits from O- to X-point, while (d) the deposition transition from X- to O-point has oppositve effect.}
\label{fig:modeAmpAndECCD}
\end{figure*}

A coherent average of the mode amplitude over 25 periods was performed for each shot, shown on the right subpanels in figure \ref{fig:modeAmpAndECCD}.
The thin dotted black lines represent one standard deviation away from the mean mode amplitude.
This averaging process reduces the noise in the signal for the following analysis.
Given the periodic nature of this behaviour, the 20~Hz frequency used in entrainment and synchronized deposition was adopted to fit a sinusoid (red) to the averaged mode amplitude (black).
This single-frequency fitting gave a \(R^2\) between 0.974 and 0.981 for the four shots, while fitting the 2nd and 3rd harmonics\textemdash with normalized amplitudes up to 0.13 and 0.04 respectively\textemdash only improved the \(R^2\) to about 0.995.
Higher harmonics were found to be progressively smaller, thus analysis proceeded with the fundamental 20~Hz fit.

The mode amplitude response to the driven current was found to be against the expectation of O-point deposition suppressing the mode.
It was actually observed that the mode amplitude had increased when the O-point was believed to be in phase with the deposition location (figure \ref{fig:modeAmpAndECCD}(a)). 
Similarly, current nominally driven in the X-point was correlated with a decrease in mode amplitude (figure \ref{fig:modeAmpAndECCD}(b)).
These peculiar results were further corroborated by two additional shots where the gyrotrons were turned on half-way between the island O- and X-points, as in figure \ref{fig:modeAmpAndECCD}(c) and (d).

\subsection{Effect of radial misalignment}
\begin{figure}[t]
\centering
\includegraphics[scale=1.1,trim={0cm 0cm 0 0cm},clip]{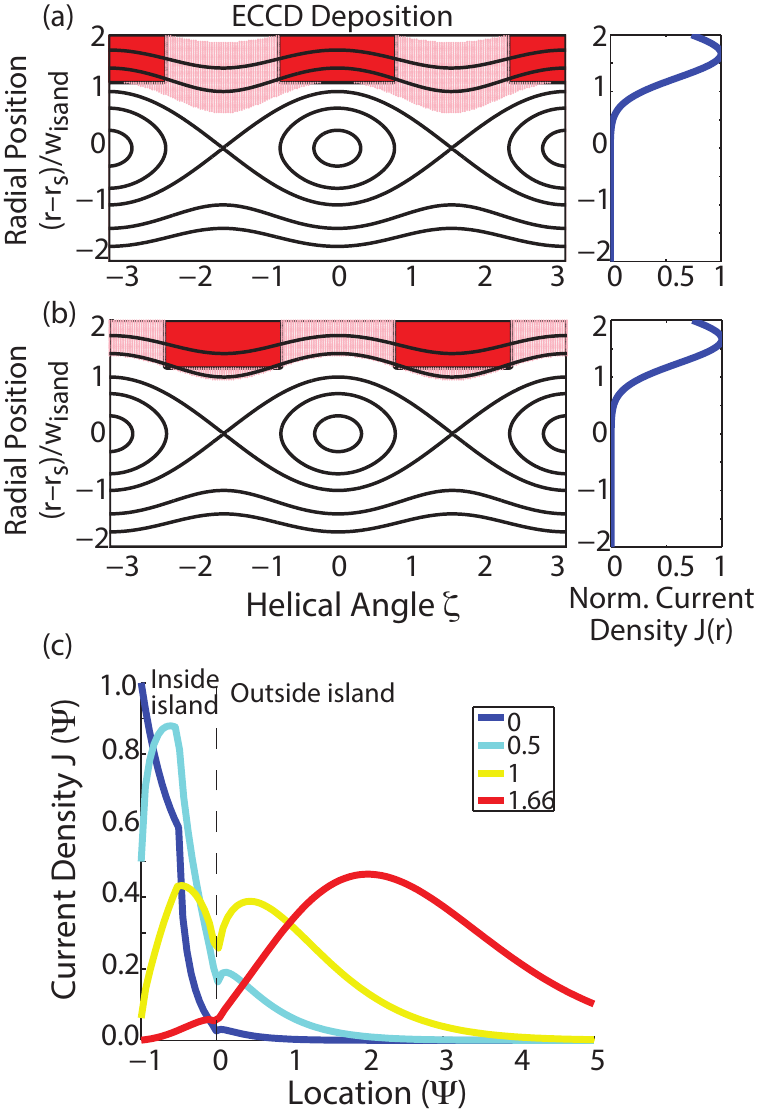}
\caption{The radial and toroidal deposition location of ECCD in shots (a) 166566 and (b) 166567, based on post-shot analysis from EFIT (mode location), TORAY-GA (ECCD location), and magnetics (island width). The solid box shows the locations near the island where EC power was deposited, which spreads over the flux surface due to parallel transport, as depicted by the shaded region. (c) shows the current drive distribution in terms of perturbed flux surface for 50\% duty cycle deposition centered on the O-point toroidally, with the color axis representing normalized radial misalignment \(|r-r_s|/w_{ECCD}\). }
\label{fig:OpointMisaligned}
\end{figure}

Upon closer analysis, it was found that the radial location of ECCD deposition, \(r\), differed from the rational surface location \(r_s\) by an amount \(|r-r_s|/w_{ECCD} = 1.66\), where \(w_{ECCD}\) is the ECCD deposition full width at half maximum (FWHM).
This was due to a poor between-shot equilibrium reconstruct based on magnetics alone and gyrotron mirrors that could only be aimed in feedforward, lacking real-time tracking of the radial position of the rational surface.
The current was driven outside of the island (solid color box in figure \ref{fig:OpointMisaligned}(a)) and redistributed, due to parallel transport, over the "intercepted" flux-surfaces, external to the island (shaded lighter color). 
X-point phasing yields figure \ref{fig:OpointMisaligned}(b), where the current is driven in flux-surfaces farther away from the island separatrix. 

When modulated ECCD is entirely deposited outside the island, it does not result in a helical current-{\em filament} that can compensate for the bootstrap current-deficit in the island. 
Rather, it results in a current-{\em sheet} or toroidal annulus being driven in a range of minor radii, but at all poloidal locations. 
At most, the current-sheet will be kink-deformed on the flux-surfaces, which are rippled due to proximity with the island. 
In these cases, scanning the phase of the time-modulation does not affect the helical phase of ECCD deposition relative to the island O-point. 
Rather, it is equivalent to {\em radially} scanning the ECCD deposition relative to the island separatrix. 

A combination of different tools were used to obtain the current deposition location relative to island position plotted in figure \ref{fig:OpointMisaligned}. 
TORAY-GA \cite{matsuda1989} provided the radial deposition profile of the ECCD, fitted with a Gaussian.
The mode's radial position is given by an EFIT equilibrium reconstruction, and its toroidal phase and width are given by magnetic measurements. 
Finally, the toroidal deposition of ECCD relative to the island phase is given by known timing.

The expected current density profile for a toroidally O-point centered 50\% duty cycle is plotted in figure \ref{fig:OpointMisaligned}(c). 
This is a function of the normalized flux-surface coordinate, perturbed by the presence of the island, introduced in \cite{perkins1997}:  
\(\Psi = (r-r_s)^2/w^2 - (1+cos(m\alpha))/2\), where \(w\) is the island width, and \(\alpha = \theta - n\phi/m\) is the helical angle.
Here \(\Psi = -1\) corresponds to the center of the island, \(\Psi = 0\) to its separatrix, and \(\Psi > 0\) to the region outside the island. 
Distinction between the core and edge sides of the island is not needed in this direct current replacement model.
Zero (dark blue) to small (light blue) radial misalignment still results in the majority of the current being driven inside the island O-point as desired.
However, increased radial misalignments places a large fraction of the current either onto separatrix (vertical dashed line), which is brought near the island X-point by parallel transport, or outside of the island altogether.
Current driven at these locations reinforces the initial perturbation, causing the mode to grow rather than be suppressed.
This reversal of effectiveness has been previously predicted by \cite{perkins1997} at sufficiently large radial misalignment \(|r-r_s|/w_{ECCD}\gtrsim 0.75\).

\subsection{Comparison with theory}
The modified Rutherford equation (MRE) is given below \cite{lahaye2002}:
\begin{dmath}
1.22^{-1}\frac{\tau_R}{r}\frac{dw}{dt} = \Delta'(w)r+
\epsilon^{1/2}\left(\frac{L_q}{L_p}\right)\beta_\theta
\left[\frac{rw}{w^2+w_d^2}-\frac{rw_{pol}^2}{w^3}-K_1\left(\frac{j_{ECCD}}{j_{BS}}\right)\right]
\label{eqn:MRE}
\end{dmath}
where \(\tau_R\) is the local resistive time and \(r\) is the minor radius of the rational surface.
The right hand side includes the classical stability index \(\Delta'(w)\) and neoclassical terms for small island effects \(w_d\) and polarization \(w_{pol}\).
The last term describes the effect of electron cyclotron current drive on the growth of the island dependent on an efficiency \(K_1\) and the ratio of ECCD and bootstrap currents.

\begin{figure*}
\begin{center}
\includegraphics[scale=1.1,trim={0cm 0cm 0 0cm},clip]{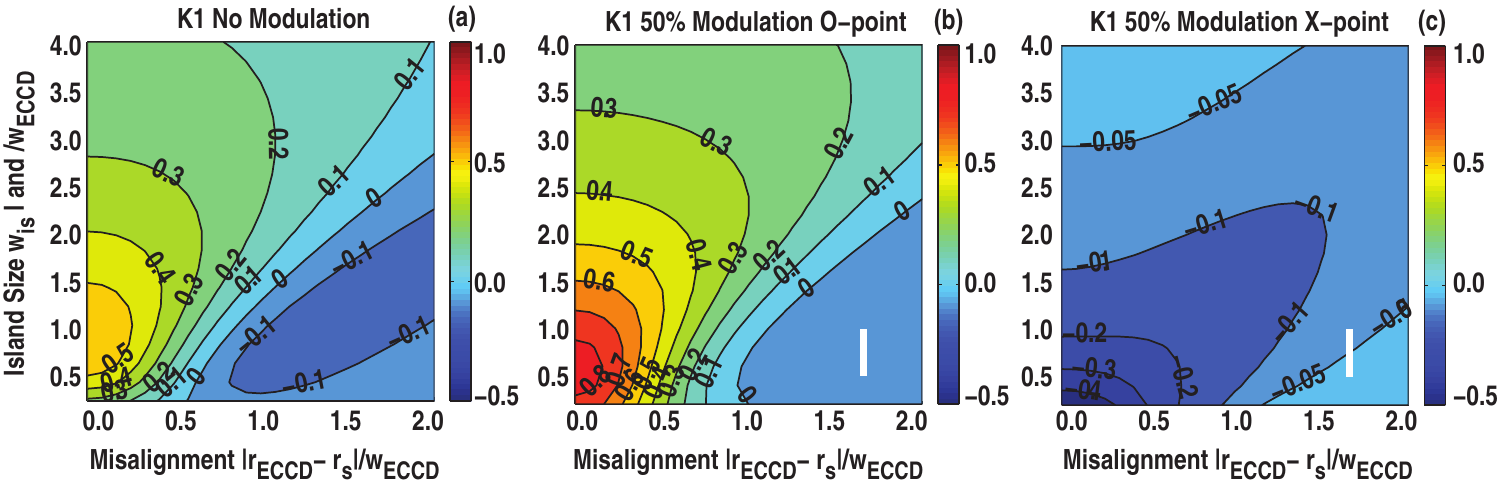}
\end{center}
\caption{Contours of ECCD deposition efficiency in terms of relative island size, radial misalignment, and duty cycle being (a) full-on, (b) 50\% centered on O-point, and (c) 50\% centered on X-point. Here, \(w_{island}\) is the island half-width, \(w_{ECCD}\) is the FWHM of the gaussian current drive profile, and misalignment is the absolute difference between peak CD location and the \(q=2\) surface. The white regions in panels (b) and (c) mark approximately the experimental values for shot 166566 and 166567 respectively.}
\label{fig:k1contours}
\end{figure*}

An early model by Perkins \cite{perkins1997} calculates the efficiency \(K_1\) of ECCD mode suppression by direct replacement of the missing bootstrap current, averaged over the modulation period. 
\begin{dmath}
K_1\left(\frac{w}{w_{ECCD}}, \frac{\Delta R}{w_{ECCD}}\right) = \int_{-1}^\infty d\Psi W(\Psi) j_{ECCD}(\Psi, w_{ECCD}, \Delta R)
\label{eqn:K1}
\end{dmath}
where the efficiency is a function of island size (affects perturbed flux \(\Psi\)), a weighting function \(W(\Psi)\), deposition width \(w_{ECCD}\), radial misalignment \(\Delta R\), as well as duty cycle and toroidal phasing.
Figure \ref{fig:k1contours} shows contours of this efficiency, where the white marks in panels (b) and (c) show estimated experimental values.
In both the cw and O-point centered modulation cases, the highest efficiency occurs for well-aligned deposition, as expected. 
O-point deposition yields higher efficiencies everywhere,  compared to cw deposition.
In the 50\% X-point modulation case (Figure \ref{fig:k1contours}(c)), a well-aligned deposition results in a moderately negative effect, which diminishes for larger misalignment or island size.

More recent work by Westerhof \cite{westerhof2016} demonstrates that the MRE can be interpreted to also include the effects of current deposited outside the island that changes the local current profile, which was neglected in this work.
Related work by Ayten and Westerhof \cite{ayten2012} further includes the effects of mode rotation, resulting in a constant cw ECCD having a time-dependent, periodic effect on stability due to being deposited at different phases relative to the island O-point.
However, the current profile can only be perturbed on an \(L/R\) timescale, where \(L\) and \(R\) are the local inductance and resistivity respectively.
In this particular experiment, the \(L/R\) time of approximately 60~ms allows us to assume current profile just outside the island does not change significantly during each 25~ms half-period of the ECCD turning on or off.

De Lazzari \cite{delazzari2009} further refined the model by including the effect of local heating on the local resistivity, and thus on the local current. 
The relative strengths of this term depend on the ratio of island and deposition widths, and on local temperature, bootstrap current, and perpendicular heat conducitivity.
This was evaluated to be roughly an order of magnitude smaller than the direct current replacement effect in this experiment.
Lastly, the change in local pressure profile gradient produces another perturbed current, but is numerically evaluated to be yet another order of magnitude smaller and therefore neglected in further analysis.
Thus, the strongest contributing term of current replacement became the focus of further analysis, neglecting the other effects.
 
\begin{figure}[t]
\includegraphics[scale=1,trim={0cm 0cm 0 0cm},clip]{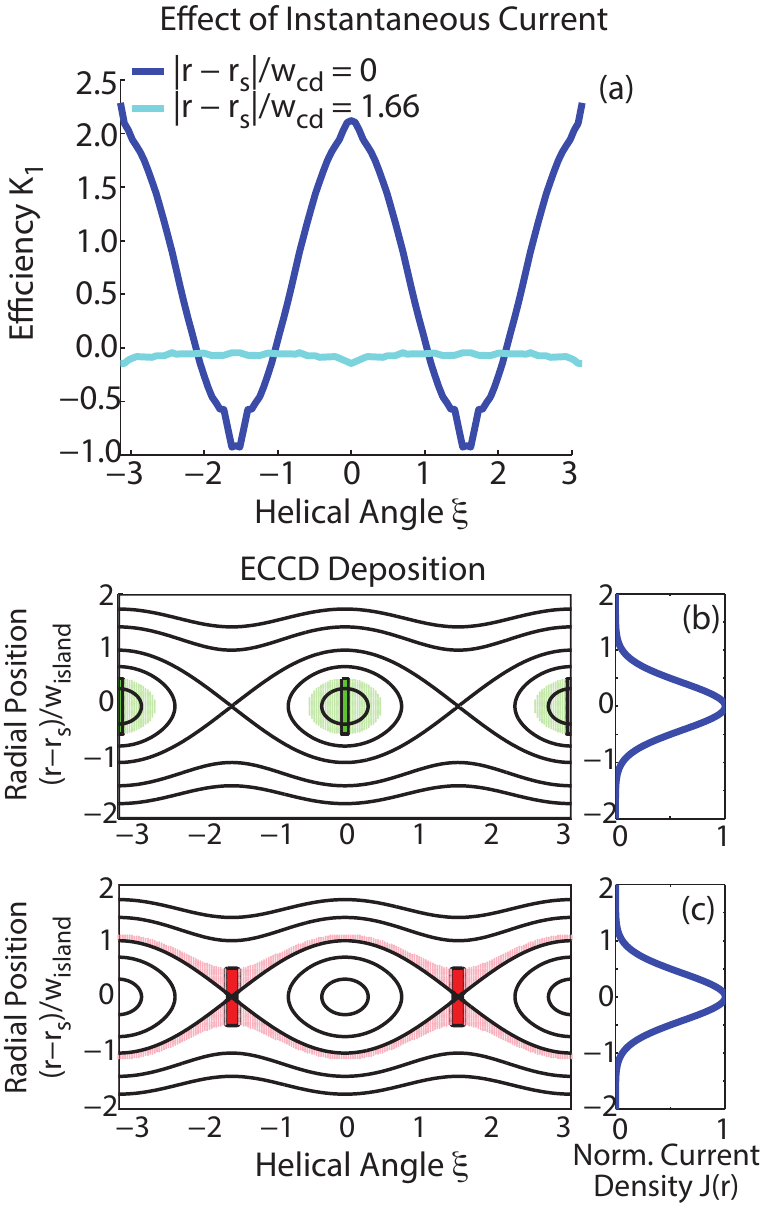}
\caption{(a) shows the instantaneous efficiency \(K_1\) for current deposited at a particular helical angle, for both radially aligned and misaligned case. (b) and (c) depicts a toroidally localized slice of current deposited in the O- and X-points of the island, respectively, where the solid box shows actual deposition location (5\% duty cycle used for clarity), and the shaded region shows the effect of current being spread over the associated flux surfaces by fast parallel transport.}
\label{fig:currentSlice}
\end{figure}

These works have been extended to calculating the change in saturated island width by using the MRE and accounting for the time-evolution of the ECCD term in the equation, due to the power being modulated and the island being entrained, changing the deposition location.
This is accomplished by evaluating the effect on the efficiency \(K_1\) of a thin slice of current deposited in a toroidally localized position, and the \(K_1\) contours in figure \ref{fig:k1contours} can be interpreted as the weighted averages over one period.
As the deposition location is moved by rotating the island, the effect of the EC suppression will change in time, resulting in a gradual shift of the saturated island width.
Figure \ref{fig:currentSlice}(b) and (c) depict a radially well-aligned blip of current in the O- and X-points of the island, and panel (a) shows the drastic difference between suppression efficiencies for radially well-aligned and misaligned depositions.
While this is consistent with mode destabilization shown in figure \ref{fig:modeAmpAndECCD}(a), panel (b) suggests that the mode is somewhat suppressed by the ECCD instead of the expected destabilization.
Further investigation is required to explain this discrepancy.
In fact, sufficiently radially misaligned deposition is expected to cause mode growth regardless of phase.
This is due to the small fraction of helical current deposited near the island X-point having a stronger negative impact on stability than the even smaller amount of current deposited near the island O-point.
This highlights a need, in future work, for robust real-time aiming of the EC mirrors towards the \(q=2\) surface in order to achieve maximum efficiency.

\section{Modeling of feedback control on ITER}
\label{sec:ITERsimulation}

The predictive model for mode dynamics was adapted to ITER to estimate the entrainment capabilities  for 2/1 islands.
In its present design, ITER will have two sets of 3D coils: the internal three rows of nine ELM control coils \cite{neumeyer2011} and the external three rows of six correction coils \cite{foussat2010}.
Continuing the phasor representation used by Olofsson \cite{olofsson2016}, each set of coils was modeled by an amplitude and phase appropriate for the applied RMP, optimized for maximum coupling (in vacuum) to the 2/1 island.
For now, the simulation assumed that the power supplies used for these 3D coils are capable of delivering the necessary currents at low frequencies (< 20~Hz).
ITER will also have a beryllium first wall and two layers of stainless steel vacuum vessel walls \cite{ioki1998}, treated here as the source of the braking torque on the island.
Using a thin wall approximation, they are reduced to a simple time constant that affects the mode dynamics and applied RMP penetration.
Other factors such as test blanket modules and divertor were not included at this stage.

\begin{figure}[t]
\centering
\subfigure{
\includegraphics[scale=0.4,trim={0cm 0cm 0 0},clip]{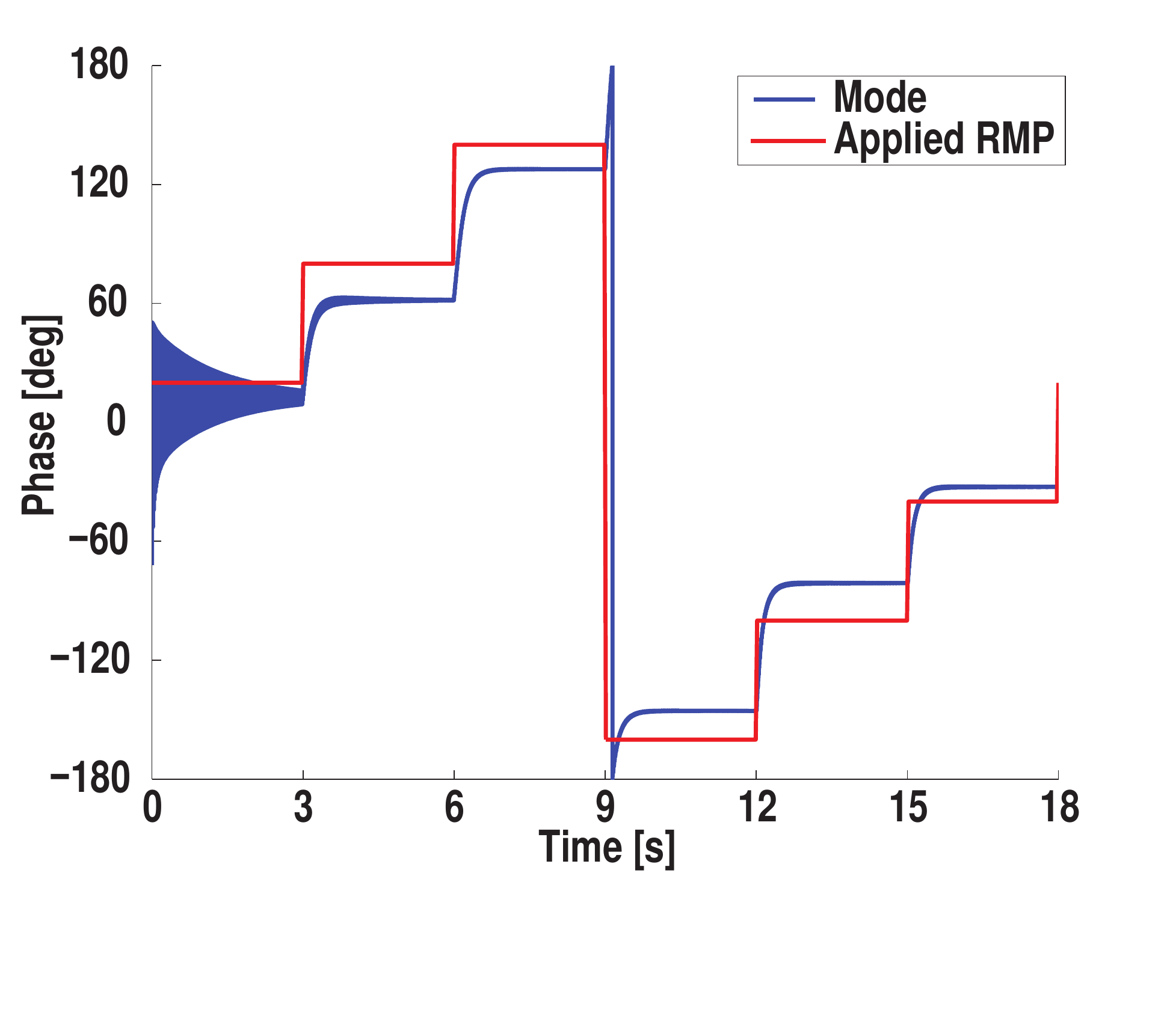}
}
\caption{Simulation for ITER of a 5~cm island (blue) following a 10~kA/turn static RMP (red), advancing in steps of 60\(^\circ\) every 3 seconds. There is an offset at each time due to the residual error field of 1~G. The initial oscillations in mode phase are simply an artifact of the magnetic field from the mode trying to penetrate the wall, which was initialized at zero. } 
\label{fig:ITERstatic}
\end{figure}

\subsection{Fixed phase}
A basic step reference, feedforward simulation was performed as a first check, using a fixed-width 5~cm island, with peak current of 10~kA/turn driven in the external coils as the RMP, and a 1~G residual error field.
The phase of the RMP is advanced by 60\(^\circ\) every 3 seconds.
Figure \ref{fig:ITERstatic} shows the mode aligning to the RMP as expected, with an offset caused by the error field.

\begin{figure}[t]
\centering
\includegraphics[scale=0.48,trim={0cm 0cm 0 0},clip]{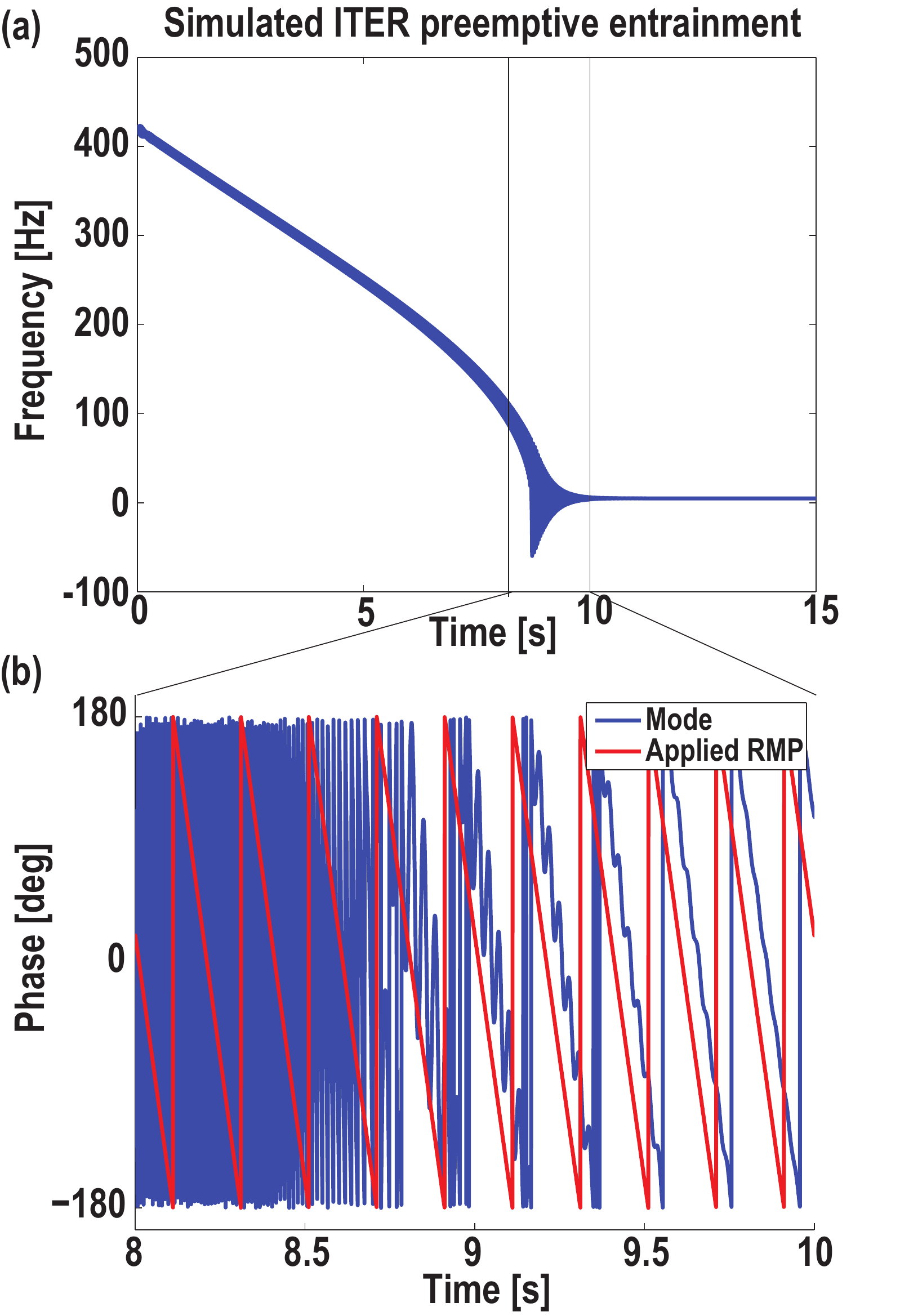}
\caption{Simulated preemptive feedforward entrainment at 5~Hz of a 5~cm island in ITER, with the rotating 10~kA RMP in the external coils turned on from the beginning. (a) shows the rotation frequency decrease and settle to 5~Hz, (b) gives the details of mode phase behaviour around locking. } 
\label{fig:ITERpreemptive}
\end{figure}

\subsection{Entrainment}
Similar to section \ref{sec:preemptive}, feedforward preemptive entrainment was simulated using the same conditions as above, except now with a rotating RMP.
Figure \ref{fig:ITERpreemptive} shows the mode slowing down and locking to the 5~Hz entraining RMP.
As entrainment is approached, the phase of the mode follows that of the RMP with some oscillations caused by the induced currents in the wall.
After these currents have fully decayed, smooth entrainment is observed for \(t \gtrsim\) 10 s.
Equation \ref{eqn:simpleEoM} was used in both of the time domain simulations, which includes the effects of error field and slow decay of shielding currents in the wall.

\begin{figure}[t]
\centering
\subfigure{
\includegraphics[scale=0.41,trim={0.1cm 0.2cm 0 0},clip]{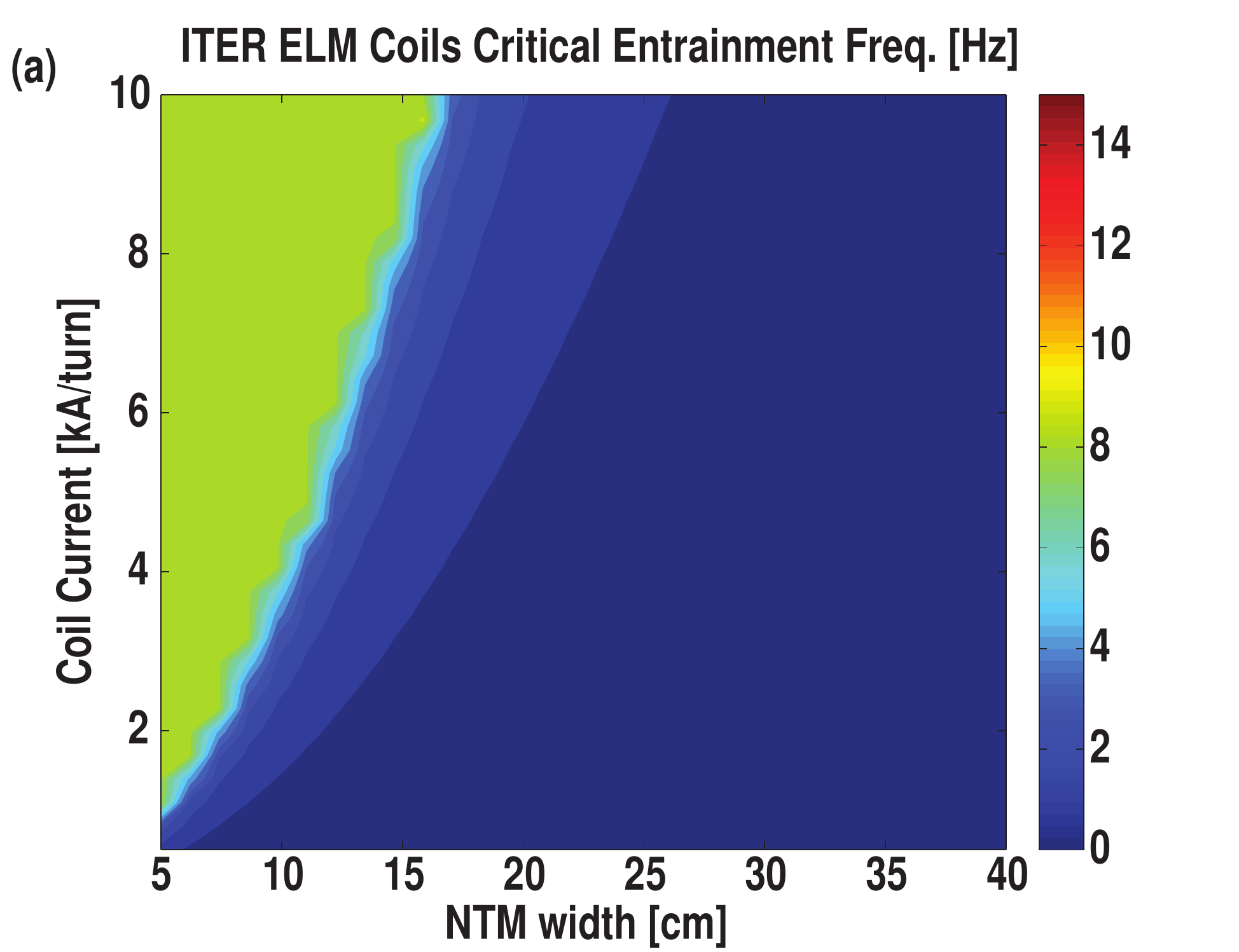}
}
\subfigure{
\includegraphics[scale=0.42,trim={0.1cm 0.2cm 0 0},clip]{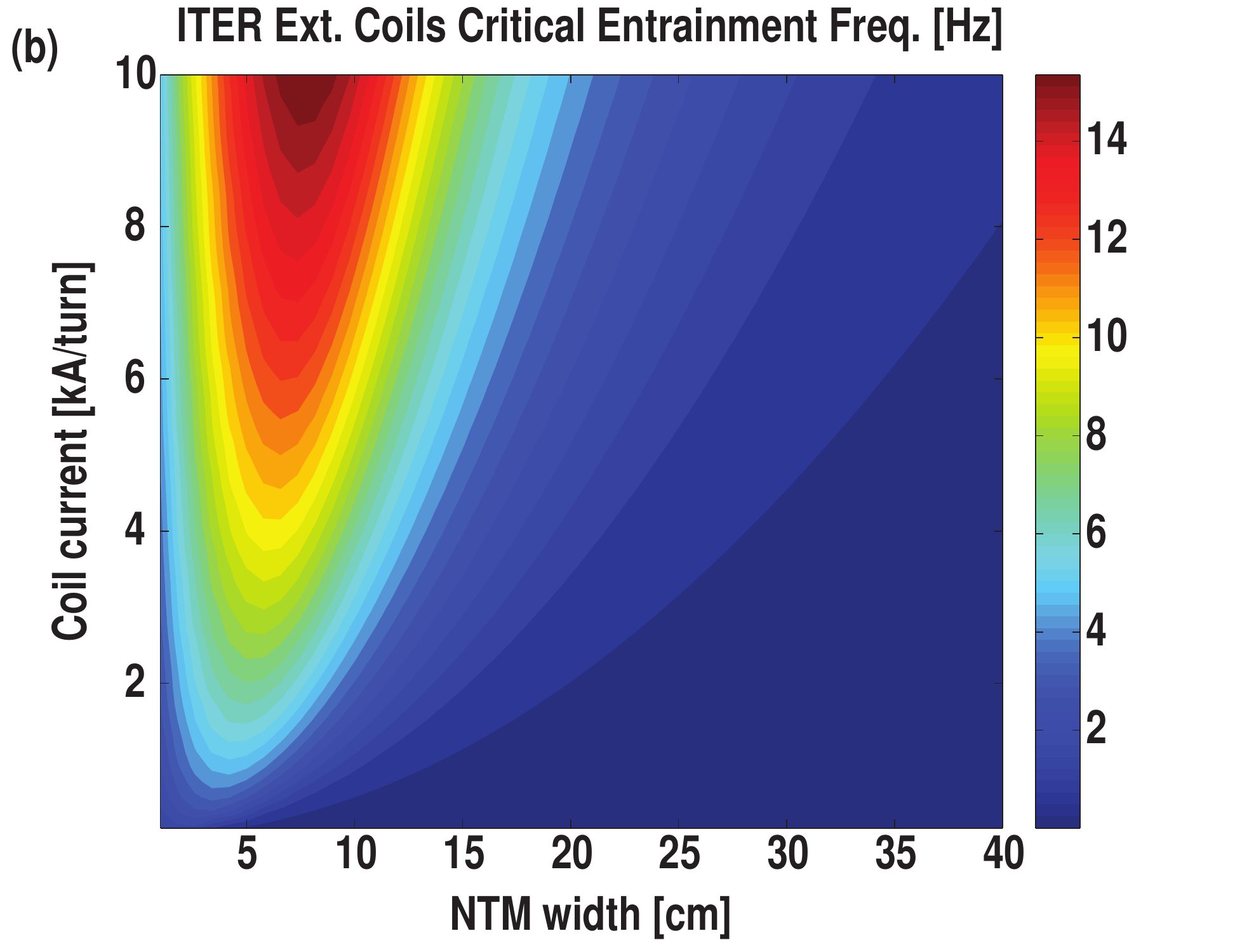}
}
\caption{Predicted maximum stable entrainment frequency contours for ITER using (a) internal ELM coils or (b) external correction coils. } 
\label{fig:ITERFreqContours}
\end{figure}

To predict the maximum entrainment frequencies possible in ITER, we searched for a steady-state torque-balance between the wall drag and the applied RMP, neglecting error fields and possible time history effects.
Equation \ref{eqn:smoothEntrain} was used in this simplified scenario to calculate the critical entrainment frequency, dependent on island width and applied current.
Figure \ref{fig:ITERFreqContours} suggests that small 2/1 islands can be entrained in the sub-10~Hz range.
Curiously, although an externally applied RMP torque must penetrate through the walls in order to affect the mode, the higher number of turns in the external coils balances out the disadvantage in location.
As a result, external coils exhibit entrainment capabilities similar to the internal coils, based on otherwise identical simulation parameters.

\section{Summary and conclusions}
\label{sec:conclusion}

Following previous work that simulated the dynamics of magnetic islands interacting with applied resonant magnetic perturbations (RMPs), error fields and conducting wall on DIII-D, a feedback controller was developed to control the phase of 2/1 islands.
Preemptive feedforward entrainment of these modes was experimentally demonstrated at DIII-D and presented here.
However, we also modeled and experimentally confirmed that feedback control enables more uniform, smoother entrainment.
Initial testing of this controller shows promising results in prescribing fixed phase to the mode, as quantified by a low tracking error.
Entrainment studies using proportional-only capabilities resulted in a higher tracking error than expected, depending on the proportional gain used and requested frequency. 
Simulation suggests larger island size or radial position also increases the tracking error.

Stable low frequency (20~Hz) entrainment was achieved in this experiment. 
Future work includes improving the controller to operate at rotation frequencies higher than the inverse wall-time, about 50~Hz at DIII-D.
This is expected to provide additional benefits for confinement, stability, and disruption avoidance.

Modulated electron cyclotron current drive (ECCD) was deposited in synchronization with the entrained, rotating mode for a detailed study of how it affects island stability, as a function of the phase relative to the island O-point.
At first sight, the relation seemed counter-intuitive, with O- and X-point deposition respectively destabilizing and slightly stabilizing the mode.
The deposition was radially misaligned, resulting in a large fraction of current being driven outside the island separatrix. 
Theoretical considerations suggest that this should be destabilizing, when the ECCD is modulated with O-point phasing, consistent with experimental observations. However, X-point phasing should also be destabilizing, albeit to a smaller degree. 
This is not consistent with the observations presented, and will be the subject of further theoretical work.
A well-aligned deposition in both radial and toroidal position would offer the most effective suppression, reducing the required input power.
The effect of modulated, misaligned heating on the temperature and pressure profiles are secondary corrections not included here, and will be investigated in future work.

The code used to model mode behaviour was adapted to ITER, where static and rotating RMPs are predicted to be able to control the island's toroidal phase.
The model also predicts a capability to entrain small (that is, recently formed, not yet saturated) islands in the sub-10~Hz range.
However, other possible effects, such as the plasma response to the applied RMP, are not studied here and are left as future work.

Applied 3D magnetic fields have long been available as a tool in studying MHD stability.
This paper presents the first instance of direct feedback phase control of 2/1 islands in real time, which can be used to prevent locking and its associated risks or to position the island favourably in order to assist its suppression by ECCD or its characterization by toroidally sparse diagnostics.

\section*{Acknowledgements}
The authors would like to thank R. Prater and X. Chen for their help in the feedforward experiment, as well as the DIII-D team for making the experiment possible. This work was realized under DOE Grants DE-SC0008520, DE-SC0016372, and DE-FC02-04ER54698.

DIII-D data shown in this paper can be obtained in digital format by following the links at https://fusion.gat.com/global/D3D_DMP.


\begin{thebibliography}{10}

\bibitem{lahaye2006}
R.J.~La Haye.
\newblock Neoclassical tearing modes and their control.
\newblock {\em Phys. Plasmas}, 13(5):055501, 2006.

\bibitem{nave1990}
M.F.F. Nave and J.A. Wesson.
\newblock Mode locking in tokamaks.
\newblock {\em Nucl. Fusion}, 30:2575, 1990.

\bibitem{chang1990}
Z~Chang and JD~Callen.
\newblock Global energy confinement degradation due to macroscopic phenomena in
  tokamaks.
\newblock {\em Nuclear Fusion}, 30(2):219, 1990.

\bibitem{morris1992}
A.W. Morris.
\newblock Mhd instability control, disruptions, and error fields in tokamaks.
\newblock {\em Plasma Phys. Control. Fusion}, 34:1871, 1992.

\bibitem{lahaye2000}
RJ~La~Haye, RJ~Buttery, S~Guenter, GTA Huysmans, M~Maraschek, and HR~Wilson.
\newblock Dimensionless scaling of the critical beta for onset of a
  neoclassical tearing mode.
\newblock {\em Physics of Plasmas}, 7(8):3349--3359, 2000.

\bibitem{volpe2015}
FA~Volpe, A~Hyatt, RJ~La~Haye, MJ~Lanctot, J~Lohr, R~Prater, EJ~Strait, and
  A~Welander.
\newblock Avoiding tokamak disruptions by applying static magnetic fields that
  align locked modes with stabilizing wave-driven currents.
\newblock {\em Physical review letters}, 115(17):175002, 2015.

\bibitem{gantenbein2000}
G~Gantenbein, H~Zohm, G~Giruzzi, S~G{\"u}nter, F~Leuterer, M~Maraschek,
  J~Meskat, Q~Yu, ASDEX~Upgrade Team, et~al.
\newblock Complete suppression of neoclassical tearing modes with current drive
  at the electron-cyclotron-resonance frequency in asdex upgrade tokamak.
\newblock {\em Physical Review Letters}, 85(6):1242, 2000.

\bibitem{hennen2010}
BA~Hennen, E~Westerhof, PWJM Nuij, JW~Oosterbeek, MR~de~Baar, WA~Bongers,
  A~B{\"u}rger, DJ~Thoen, M~Steinbuch, TEXTOR team, et~al.
\newblock Real-time control of tearing modes using a line-of-sight electron
  cyclotron emission diagnostic.
\newblock {\em Plasma Physics and Controlled Fusion}, 52(10):104006, 2010.

\bibitem{prater2004}
Ronald Prater.
\newblock Heating and current drive by electron cyclotron waves.
\newblock {\em Physics of Plasmas}, 11(5):2349--2376, 2004.

\bibitem{fitzpatrick1991}
R~Fitzpatrick and TC~Hender.
\newblock The interaction of resonant magnetic perturbations with rotating
  plasmas.
\newblock {\em Physics of Fluids B: Plasma Physics}, 3(3):644--673, 1991.

\bibitem{morris1990}
AW~Morris, TC~Hender, J~Hugill, PS~Haynes, PC~Johnson, B~Lloyd, DC~Robinson,
  C~Silvester, S~Arshad, and GM~Fishpool.
\newblock Feedback stabilization of disruption precursors in a tokamak.
\newblock {\em Physical review letters}, 64(11):1254, 1990.

\bibitem{liang2007}
Y~Liang, HR~Koslowski, A~Kr{\"a}mer-Flecken, O~Zimmermann,
  K~L{\"o}wenbr{\"u}ck, G~Bertschinger, RC~Wolf, et~al.
\newblock Observations of secondary structures after collapse events occurring
  at the q= 2 magnetic surface in the textor tokamak.
\newblock {\em Nuclear Fusion}, 47(9):L21, 2007.

\bibitem{koslowski2006}
HR~Koslowski, E~Westerhof, M~De~Bock, I~Classen, R~Jaspers, Y~Kikuchi,
  A~Kr{\"a}mer-Flecken, A~Lazaros, Y~Liang, K~L{\"o}wenbr{\"u}ck, et~al.
\newblock Tearing mode physics studies applying the dynamic ergodic divertor on
  textor.
\newblock {\em Plasma physics and controlled fusion}, 48(12B):B53, 2006.

\bibitem{volpe2009}
FA~Volpe, ME~Austin, RJ~La~Haye, J~Lohr, R~Prater, EJ~Strait, and AS~Welander.
\newblock Advanced techniques for neoclassical tearing mode control in diii-d
  a.
\newblock {\em Physics of Plasmas}, 16(10):102502, 2009.

\bibitem{shiraki2014}
D~Shiraki, RJ~La~Haye, NC~Logan, EJ~Strait, and FA~Volpe.
\newblock Error field detection in diii-d by magnetic steering of locked modes.
\newblock {\em Nuclear Fusion}, 54(3):033006, 2014.

\bibitem{okabayashi2014}
M~Okabayashi.
\newblock {\em General Atomics Report}, pages GA--A27926, 2014.

\bibitem{okabayashi2017}
M.~Okabayashi, P.~Zanca, E.J. Strait, A.M. Garofalo, J.M. Hanson, Y.~In,
  R.J.~La Haye, L.~Marrelli, P.~Martin, R.~Paccagnella, C.~Paz-Soldan,
  P.~Piovesan, C.~Piron, L.~Piron, D.~Shiraki, F.A. Volpe, The DIII-D, and RFX
  mod Teams.
\newblock Avoidance of tearing mode locking with electro-magnetic torque
  introduced by feedback-based mode rotation control in diii-d and rfx-mod.
\newblock {\em Nucl. Fusion}, 57(1):016035, 2017.

\bibitem{volpe2014}
F~Volpe.
\newblock Using 3d fields to control islands, aid eccd-stabilization and
  measure error-fields in diii-d, 2014.

\bibitem{devries1996}
PC~De~Vries, G~Waidmann, AJH Donn{\'e}, and FC~Sch{\"u}ller.
\newblock Mhd-mode stabilization by plasma rotation in textor.
\newblock {\em Plasma physics and controlled fusion}, 38(4):467, 1996.

\bibitem{maraschek2007}
M~Maraschek, G~Gantenbein, Q~Yu, H~Zohm, S~G{\"u}nter, F~Leuterer, A~Manini,
  ECRH Group, ASDEX~Upgrade Team, et~al.
\newblock Enhancement of the stabilization efficiency of a neoclassical
  magnetic island by modulated electron cyclotron current drive in the asdex
  upgrade tokamak.
\newblock {\em Physical review letters}, 98(2):025005, 2007.

\bibitem{olofsson2016}
KEJ Olofsson, W~Choi, DA~Humphreys, RJ~La~Haye, D~Shiraki, R~Sweeney, FA~Volpe,
  and AS~Welander.
\newblock Electromechanical modelling and design for phase control of locked
  modes in the diii-d tokamak.
\newblock {\em Plasma Physics and Controlled Fusion}, 58(4):045008, 2016.

\bibitem{brennan2014}
DP~Brennan and JM~Finn.
\newblock Control of linear modes in cylindrical resistive magnetohydrodynamics
  with a resistive wall, plasma rotation, and complex gain.
\newblock {\em Physics of Plasmas}, 21(10):102507, 2014.

\bibitem{chen1990}
XL~Chen and PJ~Morrison.
\newblock The effect of viscosity on the resistive tearing mode with the
  presence of shear flow.
\newblock {\em Physics of Fluids B: Plasma Physics}, 2(11):2575--2580, 1990.

\bibitem{buttery2008}
RJ~Buttery, RJ~La~Haye, P~Gohil, GL~Jackson, H~Reimerdes, and EJ~Strait.
\newblock The influence of rotation on the $\beta$ n threshold for the 2/ 1
  neoclassical tearing mode in diii-d a.
\newblock {\em Physics of Plasmas}, 15(5):056115, 2008.

\bibitem{lahaye2009}
RJ~La~Haye and RJ~Buttery.
\newblock The stabilizing effect of flow shear on m/n= 3/2 magnetic island
  width in diii-d.
\newblock {\em Physics of Plasmas}, 16(2):022107, 2009.

\bibitem{austin2003}
ME~Austin and J~Lohr.
\newblock Electron cyclotron emission radiometer upgrade on the diii-d tokamak.
\newblock {\em Review of scientific instruments}, 74(3):1457--1459, 2003.

\bibitem{strait2006}
EJ~Strait.
\newblock Magnetic diagnostic system of the diii-d tokamak.
\newblock {\em Review of scientific instruments}, 77(2):023502, 2006.

\bibitem{king2014}
Joshua~D King, Edward~J Strait, Rejean~L Boivin, Doug Taussig, Matthias~G
  Watkins, Jeremy~M Hanson, Nikolas~C Logan, Carlos Paz-Soldan, David~C Pace,
  Daisuke Shiraki, et~al.
\newblock An upgrade of the magnetic diagnostic system of the diii-d tokamak
  for non-axisymmetric measurements.
\newblock {\em Review of Scientific Instruments}, 85(8):083503, 2014.

\bibitem{hanson2016}
JM~Hanson, J~Bialek, F~Turco, J~King, GA~Navratil, EJ~Strait, and A~Turnbull.
\newblock Validation of conducting wall models using magnetic measurements.
\newblock {\em Nuclear Fusion}, 56(10):106022, 2016.

\bibitem{ogataControlText}
Katsuhiko Ogata.
\newblock {\em Modern Control Engineering}.
\newblock 2010.

\bibitem{felici2015}
Federico Felici and Tom Oomen.
\newblock Enhancing current density profile control in tokamak experiments
  using iterative learning control.
\newblock In {\em Decision and Control (CDC), 2015 IEEE 54th Annual Conference
  on}, pages 5370--5377. IEEE, 2015.

\bibitem{ravensbergen2017}
Timo Ravensbergen, Peter~C de~Vries, Federico Felici, Thomas~Cornelis Blanken,
  Remy Nouailletas, and L~Zabeo.
\newblock Density control in iter: an iterative learning control and robust
  control approach.
\newblock {\em Nuclear Fusion}, 58(1):016048, 2017.

\bibitem{lohr2005}
John Lohr, YA~Gorelov, K~Kajiwara, Dan Ponce, RW~Callis, JL~Doane, RL~Ellis,
  HJ~Grunloh, CP~Moeller, J~Peavey, et~al.
\newblock The electron cyclotron resonant heating system on the diii-d tokamak.
\newblock {\em Fusion science and technology}, 48(2):1226--1237, 2005.

\bibitem{matsuda1989}
Kyoko Matsuda.
\newblock Ray tracing study of the electron cyclotron current drive in diii-d
  using 60 ghz.
\newblock {\em IEEE transactions on plasma science}, 17(1):6--11, 1989.

\bibitem{perkins1997}
FW~Perkins, RW~Harvey, M~Makowski, and MN~Rosenbluth.
\newblock Prospects for electron cyclotron current drive stabilization of
  neoclassical tearing modes in iter.
\newblock {\em 24th EPS Conference Proceedings}, page 1017, 1997.

\bibitem{lahaye2002}
RJ~La~Haye, S~G{\"u}nter, DA~Humphreys, J~Lohr, TC~Luce, ME~Maraschek,
  CC~Petty, R~Prater, JT~Scoville, and EJ~Strait.
\newblock Control of neoclassical tearing modes in diii--d.
\newblock {\em Physics of Plasmas}, 9(5):2051--2060, 2002.

\bibitem{westerhof2016}
E~Westerhof, HJ~de~Blank, and J~Pratt.
\newblock New insights into the generalized rutherford equation for nonlinear
  neoclassical tearing mode growth from 2d reduced mhd simulations.
\newblock {\em Nuclear Fusion}, 56(3):036016, 2016.

\bibitem{ayten2012}
B~Ayten and E~Westerhof.
\newblock Consequences of plasma rotation for neoclassical tearing mode
  suppression by electron cyclotron current drive.
\newblock {\em Physics of Plasmas}, 19(9):092506, 2012.

\bibitem{delazzari2009}
D~De~Lazzari and E~Westerhof.
\newblock On the merits of heating and current drive for tearing mode
  stabilization.
\newblock {\em Nuclear Fusion}, 49(7):075002, 2009.

\bibitem{neumeyer2011}
C.~Neumeyer.
\newblock Design of the iter in-vessel coils.
\newblock {\em Fusion Sci. Technol.}, 60(1):95--99, 2011.

\bibitem{foussat2010}
A~Foussat, P~Libeyre, N~Mitchell, Y~Gribov, CTJ Jong, D~Bessette, R~Gallix,
  P~Bauer, and A~Sahu.
\newblock Overview of the iter correction coils design.
\newblock {\em IEEE Transactions on Applied Superconductivity}, 20(3):402--406,
  2010.

\bibitem{ioki1998}
K~Ioki, V~Barabash, A~Cardella, F~Elio, Y~Gohar, G~Janeschitz, G~Johnson,
  G~Kalinin, D~Lousteau, M~Onozuka, et~al.
\newblock Design and material selection for iter first wall/blanket, divertor
  and vacuum vessel.
\newblock {\em Journal of nuclear materials}, 258:74--84, 1998.

\end{thebibliography}

\end{document}